\documentclass[pra,aps,twocolumn,showpacs,superscriptaddress,nofootinbib]{revtex4-1}
\usepackage[utf8]{inputenc}
\usepackage[T1]{fontenc} 
\usepackage{amssymb}
\usepackage{amsmath}
\usepackage{color}
\usepackage{graphicx}
\usepackage{hyperref}
\usepackage{microtype}
\usepackage{dcolumn}
\usepackage{bm}
\usepackage{epsfig}

\usepackage{amsfonts}
\usepackage{graphics}
\usepackage{bbold}
\usepackage{braket}
\usepackage{empheq}

\newcommand{\titlestr}{Ultraefficient Coupling of a Quantum Emitter to the \\ Tunable Guided Plasmons of a Carbon Nanotube}

\hypersetup{colorlinks,hypertexnames=true,pdftitle={\titlestr},linkcolor=blue,citecolor=blue,urlcolor=blue}

\def\kp{{k_{\rm p}}}  \def\rb{{\bf r}}  \def\pb{{\bf p}}  \def\ii{{\rm i}}
\def\WW{W}
\def\ii{{\rm i}}  \def\ee{{\rm e}}  \def\jb{{\bf j}}  \def\Eb{{\bf E}}  \def\rb{{\bf r}}

\newcommand*\widefbox[1]{\fbox{\hspace{2em}#1\hspace{2em}}}
\def\ii{{\rm i}}  \def\ee{{\rm e}}  \def\jb{{\bf j}}  \def\Eb{{\bf E}}  \def\rb{{\bf r}}

\def\M{\vec{M}_{n, k, k_z}}
\def\N{\vec{N}_{n, k, k_z}}

\def\Md{\vec{M}^\dagger_{n, k, k_z}}
\def\Nd{\vec{N}^\dagger_{n, k, k_z}}
\def\Mdp{\vec{M}^\dagger_{n', k', k'_z}}
\def\Ndp{\vec{N}^\dagger_{n', k', k'_z}}
\def\Gm{\overline{\overline{G}}_m}
\def\Gmone{\overline{\overline{G}}_{m1}}
\def\Gmtwo{\overline{\overline{G}}_{m2}}
\def\Ge{\overline{\overline{G}}_e}
\def\Geone{\overline{\overline{G}}_{e1}}
\def\Getwo{\overline{\overline{G}}_{e2}}
\def\vGm{\overline{\overline{G}}_{m0}}
\def\rGm{\overline{\overline{G}}_{mr}}
\def\vGe{\overline{\overline{G}}_{e0}}
\def\rGe{\overline{\overline{G}}_{er}}
\def\tGm{\overline{\overline{G}}_{mt}}
\def\tGe{\overline{\overline{G}}_{et}}
\def\alphabar{\overline{\alpha}}
\def\bjn{J_{n}}
\def\bhn{H_{n}}
\def\bjpn{{J'_{n}}}
\def\bhpn{{H'_{n}}}
\def\alphai{\bar{\alpha}}

\def\kp{{k_{\rm p}}}  \def\rb{{\bf r}}  \def\pb{{\bf p}}  \def\ii{{\rm i}}
\def\WW{W}

\begin{document}
\title{\titlestr}
\author{Luis~Mart\'{\i}n-Moreno}
\email{lmm@unizar.es}
\affiliation{Instituto de Ciencia de Materiales de Arag\'on and Departamento de F\'{\i}sica de la Materia Condensada, CSIC-Universidad de Zaragoza, E-50009, Zaragoza, Spain}
\author{F.~Javier~Garc\'{\i}a~de~Abajo}
\email{ javier.garciadeabajo@icfo.es}
\affiliation{ICFO - Institut de Ciencies Fotoniques,
The Barcelona Institute of Science and Technology, 08860 Castelldefels (Barcelona), Spain}
\affiliation{ICREA - Instituci\'o Catalana de Recerca i Estudis Avan\c{c}ats, Barcelona, Spain}
\author{Francisco~J.~Garc\'{\i}a-Vidal}
\affiliation{Departamento de F\'isica Te\'orica de la Materia Condensada
and Condensed Matter Physics Center (IFIMAC), Universidad Aut\'onoma de Madrid, E-28049 Madrid, Spain}
\affiliation{Donostia International Physics Center (DIPC), E-20018 Donostia/San Sebastian, Spain}

\date{\today}

\begin{abstract}
We show that a single quantum emitter can efficiently couple to the tunable plasmons of a highly doped single-wall carbon nanotube (SWCNT). Plasmons in these quasi-one-dimensional carbon structures exhibit deep subwavelength confinement that pushes the coupling efficiency close to $100\%$ over a very broad spectral range. This phenomenon takes place for distances and tube diameters comprising the nanometer and micrometer scales. In particular, we find a $\beta$ factor $\approx1$ for QEs placed $1-100$ nm away from SWCNTs that are just a few nanometers in diameter, while the corresponding Purcell factor exceeds $10^6$. Our finding not only holds great potential for waveguide QED, in which an efficient interaction between emitters and cavity modes is pivotal, but it also provides a way of realizing quantum strong coupling between several emitters mediated by SWCNT plasmons, which can be controlled through the large electro-optical tunability of these excitations.
\end{abstract}
\pacs{78.67.Wj,73.20.Mf}
\maketitle

\section{Introduction}

Achieving an efficient coupling between a single quantum emitter (QE) and the surface plasmons (SPs) supported by metallic nanostructures has become a popular subject of research due to its potential application to quantum-optics \cite{CSH06,Akimov07,CSD07} and sensing \cite{NE97,KLN08}. This efficient coupling lies at the heart of several surface-based ultrasensitive optical analysis techniques, which rely on the plasmon-driven enhancement of Raman scattering \cite{NE97} and infrared absorption \cite{KLN08}. Remarkably, the \emph{localized} SPs of a metal nanoparticle can enormously modify the spontaneous decay of a neighboring excited molecule \cite{Anger06,Kuhn06,AMG10}, while \emph{propagating} SPs can produce similar effects over a broadband spectral range. Reducing the dimensionality of the plasmonic structure from 2D (metal surfaces) to 1D (thin wires) enables better control over the coupling, which can be engineered to affect just a single SP \cite{Akimov07}. Additionally, the SPs of 1D geometries are well suited to act as mediators in the interaction between several QEs placed in close proximity to a plasmonic waveguide \cite{Martin-Cano10,DSF10,Gonzalez-Tudela11,Dzsotjan11}, thus suggesting the combination of these tools to design large-scale quantum-optics integrated devices, which could benefit from the plasmon robustness against environmental fluctuations to operate under ambient conditions.

The recent emergence of graphene as a plasmonic material \cite{Hanson08,Jablan09,CastroNeto} has introduced an additional knob to improve the performance of QE-SP coupling. The large electrical tunability and high degree of confinement recently measured in graphene plasmons \cite{Fei11,chen12,FRA12,Fang13,BJS13} has stimulated suggestions for their use in tunable plasmonic circuitry and metamaterials \cite{Vakil11}, as well as for the achievement of quantum strong coupling and efficient interaction with QEs \cite{Koppens11,Nikitin11} with superior performance compared with conventional plasmonic metals. Although these plasmons have been so far observed only at mid-infrared and lower frequencies, their extension towards the more technologically appealing spectral ranges of the visible and near-infrared has been argued to be attainable \cite{DeAbajo14}, particularly by reducing the size of the structures to scales of a few-nanometers, which are commensurable with existing graphene-related structures such as aromatic molecules \cite{Manjavacas13} and carbon nanotubes. In particular, nanotubes of tens of nanometers in diameter have been recently suggested as suitable elements for plasmon circuitry \cite{ISL14}. It should be noted that SWCNTs, like other carbon allotropes, exhibit UV plasmons that have been well characterized in the past \cite{STK02}. However, those plasmons are much lossier, and therefore less prone to efficiently couple to QEs, than the tunable lower-energy plasmons on which we concentrate here, which only exist in doped structures and are predicted to display similar electrical tunability as graphene.

In this paper, we show that quantum emitters can strongly couple to the electrically tunable plasmons of doped SWCNTs, reaching light-matter interaction levels that go even beyond those of planar graphene. The Purcell factor (i.e., the decay rate near the material, normalized to the decay rate in free space) is increased by nearly three orders of magnitude when reducing the dimensionality of the carbon nanostructure from 2D (graphene) to quasi 1D (nanotubes). More importantly, the coupling efficiency of the QE to the SPs supported by SWCNTs (i.e., the fraction of decay into plasmons, also known as $\beta$ factor) reaches values nearing $100\%$ over a very broad range of QE-SWCNT distances and QE/SP frequencies.

\begin{figure}[htbp]
\includegraphics[width=\columnwidth]{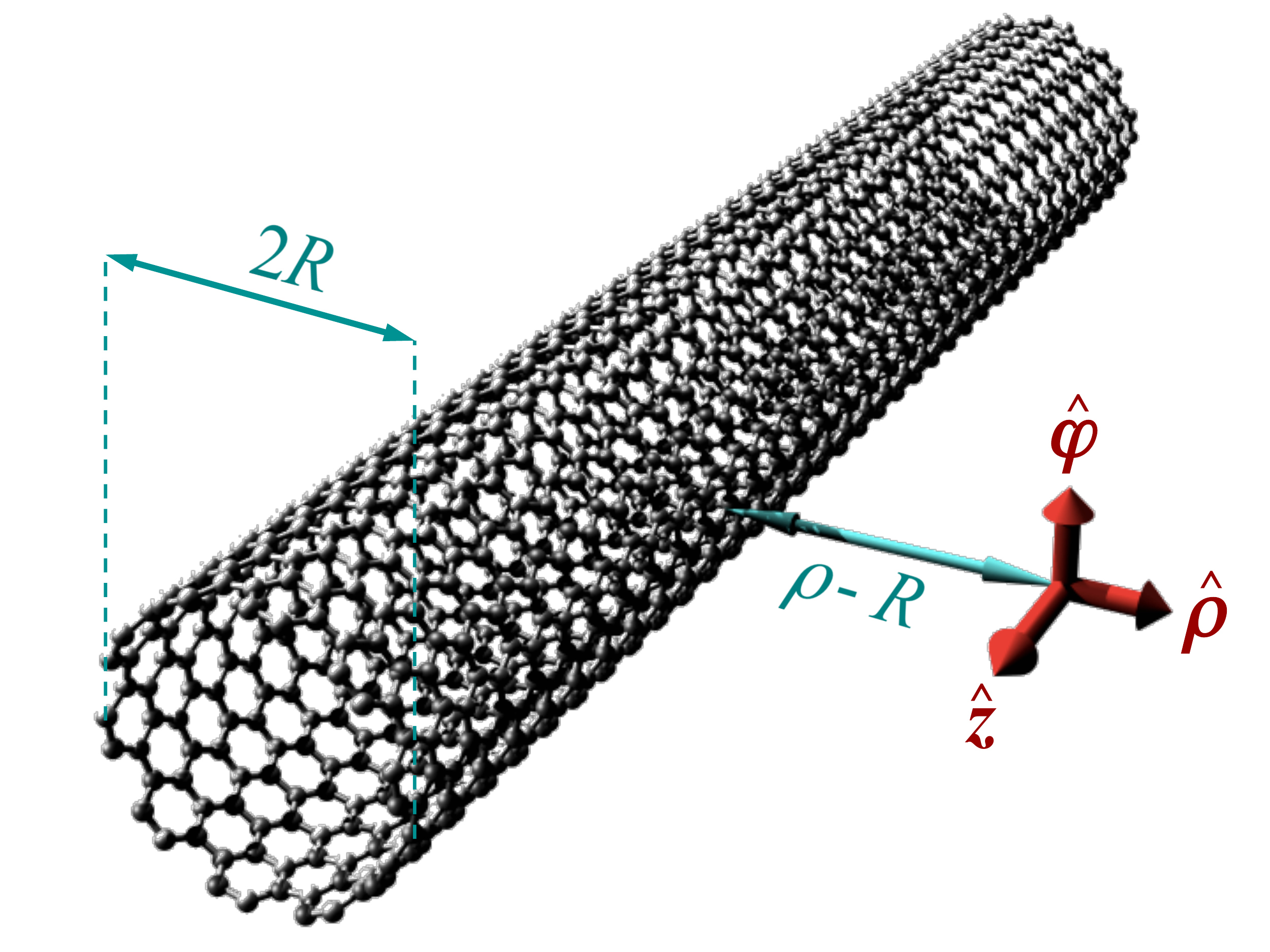}
\caption{{\bf Sketch of the system under study.} We consider a QE placed at a distance $\rho-R$ from the surface of a SWCNT of radius $R$. Three orthogonal orientations of the emitter dipole $\pb$ are considered, as shown by the red arrows.}
\label{Fig1}
\end{figure}

\section{Theoretical framework}

Figure\ \ref{Fig1} depicts the system under study: a single QE placed at a distance $\rho$ from the axis of a SWCNT of radius $R$. The emission properties of the QE are determined by its transition frequency $\omega$ and dipole moment $\pb$. We assume that $R$ is sufficiently large as to neglect curvature effects (i.e., the discreteness of the electronic bands, as well as features near the Dirac point associated with the tube chirality and finite radius). A recent study \cite{DeAbajo14} indicates that this approximation works well to describe transversal plasmons for $R>1\,$nm under the doping conditions here considered (see below), although narrower nanotubes require a more fundamental level of description \cite{GCW04,SS10}. SWCNTs have been synthesized in this size range \cite{Ma09} and their excitonic absorption bands well characterized \cite{SEW06,BPT13,LHC14}. We thus model the SWCNT as a hollow tube with the same surface conductivity $\sigma(\omega)$ as graphene doped with the considered number of charge carriers per carbon atom. This allows us to readily compare 2D graphene and 1D SWCNTs. For simplicity, we adopt the Drude model, $\sigma(\omega)=(\ii e^2E_{\rm F}/\pi \hbar^2)/(\omega+\ii/\tau)$, with realistic values of the Fermi energy $E_F=1$\,eV and the plasmon relaxation time $\tau=1$\,ps. Importantly, the relative comparison between both materials is independent of the choice of $\tau$. However, the high Purcell factor here predicted (see below) is roughly proportional to the assumed $\tau$. In this work we are most interested on situations in which both $R$ and distances $\rho$ are small compared with the light wavelength $\lambda$. Therefore, the electrostatic limit here outlined, which leads to relatively affordable analytical expressions, provides a very accurate level of description, as shown below. A comparison between results obtained with the electrostatic approach and a full electromagnetic formalism \cite{tbp} is presented in Fig.\ \ref{Fig2}, whereas both approaches give nearly identical results on the scale of Figs. \ref{Fig3} and \ref{Fig4}.

We start our analysis by considering the screened interaction $\WW(\rb,\rb',\omega)$, defined as the electric scalar potential created at the position $\rb$ by an oscillating point charge $\exp(-\ii\omega t)$ placed at $\rb'$. Reciprocity implies that $\WW(\rb,\rb',\omega)$ is symmetric with respect to the exchange of $\rb$ and $\rb'$. Also, it is convenient to decompose $\WW(\rb',\rb,\omega)=1/|\rb-\rb'|+\WW^{\rm ind}(\rb',\rb,\omega)$ as the sum of bare and induced interactions. This quantity allows us to obtain the plasmon characteristics, as well as the decay rate of a neighboring QE. Direct solution of Poisson's equation for both $\rb$ and $\rb'$ placed outside the tube yields (see Appendix)
\begin{align}
&{\WW}^{\rm ind}(\rb,\rb',\omega) \!=\! \frac{2}{\pi}\sum_{m=0}^{\infty}(2-\delta_{m0})\cos[m(\varphi-\varphi^{\prime})]
\label{W}\\
&\times\;\int_{0}^{\infty}\!\!\!\!dk\;r_m(k)\;\cos[k(z-z^{\prime})]\; K_m(k\rho)K_m(k\rho'), \nonumber
\end{align}
where we use cylindrical coordinates $\rb=(\rho,z,\varphi)$,
\begin{equation}
r_m(k)=\frac{-I_m^2(kR)\Delta_m}{1+I_m(kR)K_m(kR)\Delta_m}
\label{rm}
\end{equation}
is the reflection coefficient for cylindrical waves, $\Delta_m=(4\pi\ii\sigma/\omega R)(m^2+k^2R^2)$, and $I_m$ and $K_m$ are modified Bessel functions. The integral in Eq.\ (\ref{W}) is performed over the wave vector $k$ parallel to the axis of the nanotube, while the sum runs over components of fixed azimuthal angular momentum number $m$.

\section{Plasmon dispersion relation}

\begin{figure}[htbp]
\includegraphics[width=\columnwidth]{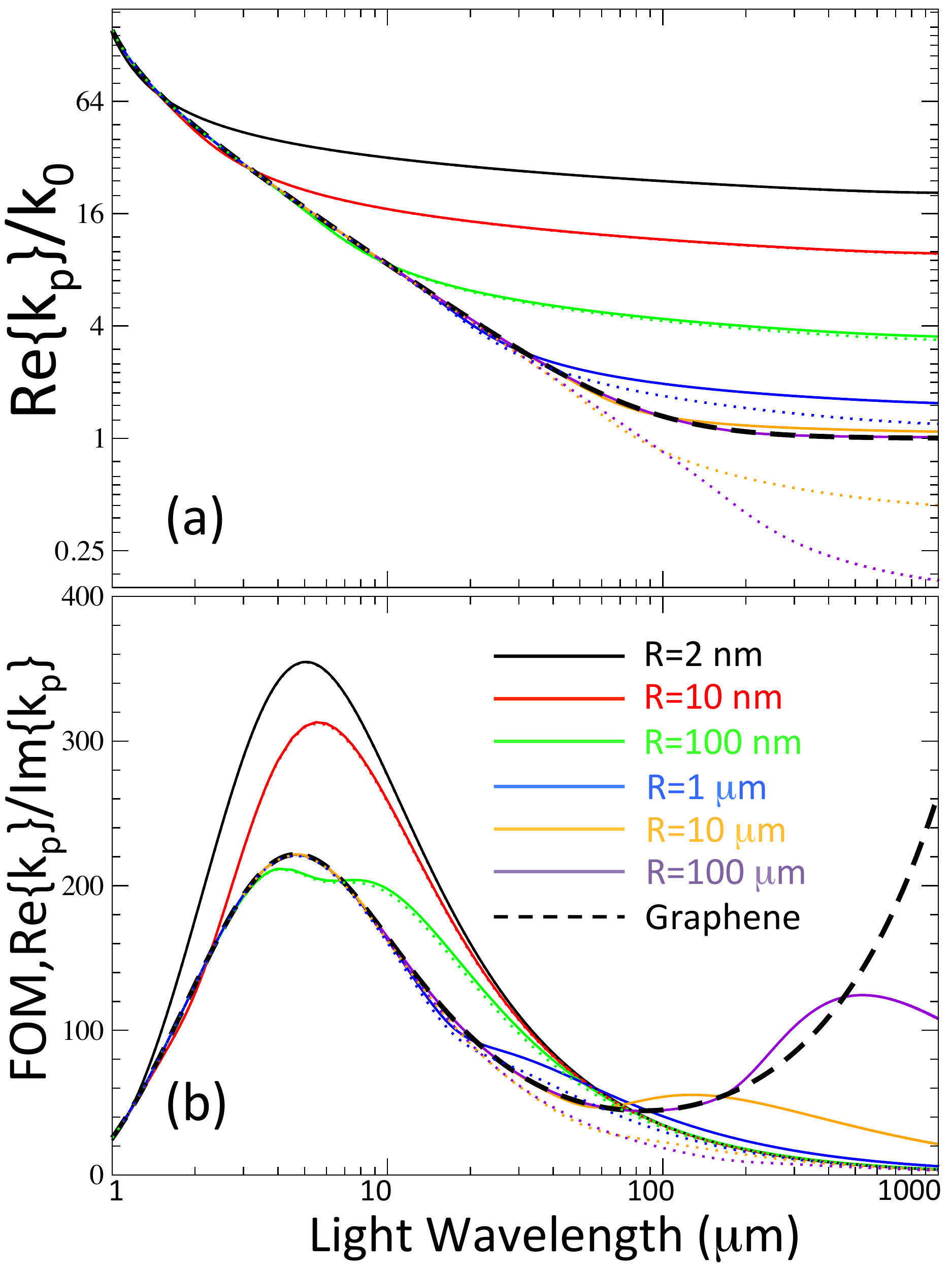}
\caption{{\bf Propagation characteristics of plasmons supported by SWCNTs and the limit towards graphene.} {\bf (a)} Real part of the plasmon wave vector $\kp$ as a function of the wavelength $\lambda$ of free-space light oscillating at the same frequency for different values of tube radius $R$. The plasmon wave vector is normalized to the light wave vector $k_0=2\pi/\lambda$. The limit of 2D graphene (dashed curve) is smoothly approach at large $R$'s. {\bf (b)} The corresponding figure of merit (FOM) ${\rm Re}\{\kp\}/{\rm Im}\{\kp\}$ of the guided plasmons studied in (a). Full electromagnetic theory (solid curves) is compared with the electrostatic limit (dotted curves) in both panels.}
\label{Fig2}
\end{figure}

Plasmon resonances are signaled by their strong response for a given external perturbation, or equivalently, by the poles of $r_m(k)$. Here we should note that sign cancellations due to the $\exp(\ii m\varphi)$ modulation of the induced charge along the azimuthal direction of the tube surface render the contribution of $m\neq0$ modes small if $\rho$ is larger than the radius $R$. We thus concentrate on the dominant $m=0$ plasmon band, whose complex wave vector $\kp$ is found as a function of frequency $\omega$ from the solution to the transcendental equation
\begin{equation}
\omega(\omega+\ii/\tau)=\frac{4e^2E_{\rm F}}{\hbar^2}\,I_0(\kp R)K_0(\kp R)\;k_{\rm p}^2R.
\nonumber
\end{equation}
This dispersion relation agrees with previous studies that focus on the $\tau \rightarrow \infty$ limit \cite{Longe93, Yannouleas96, Jiang96}. In Fig.\ \ref{Fig2}(a) we show ${\rm Re}\{\kp\}$ as a function of light wavelength $\lambda$ for SWCNTs of radius $R$ in the $2\,$nm$-100\,\mu$m range. This magnitude is normalized to the free-space light wave vector $k_0=2\pi/\lambda$, so that the plot directly illustrates the degree of spatial confinement of the plasmon, whose radial electric field is proportional to $K_1(\kp\rho)$. Whereas the case of small radius corresponds to realistic SWCNTs, we also consider very large $R$ in order to also deal with graphene coated cylinders. In this way, this figure illustrates the evolution from small tubes to planar graphene with increasing $R$. For comparison, we also plot the dispersion relation of the SPs supported by a graphene sheet, clearly showing that plasmons propagating along carbon nanotubes are more confined (larger ${\rm Re}\{\kp\}$) than those supported by graphene, and their confinement increases with decreasing $R$. This suggests that SWCNTs are better suited to produce enhanced coupling with QEs. As discussed above, the electrostatic limit yields very accurate results for nanotubes of small radius ($R<100$ nm), and also for larger radius at short wavelengths.

The ratio between the real and imaginary parts of $\kp$ is also an important magnitude, typically used as a figure of merit (FOM=${\rm Re}\{\kp\}/{\rm Im}\{\kp\}$) for evaluating the propagation characteristics of SPs. This FOM is plotted in Fig.\ \ref{Fig2}(b) as a function of $\lambda$ for different SWCNTs of different radius $R$. Remarkably, these plasmons possess a larger FOM than those of graphene for spectral and geometrical-parameter ranges in which they also exhibit tighter confinement, as noted above (see $\lambda<100\,\mu$m region). These characteristics are very beneficial for the design of efficient coupling schemes between several QEs mediated by SPs.

\section{Purcell factor}

A convenient way of assessing the strength of the QE-SP coupling consists in analyzing the rate of spontaneous emission, $\Gamma$. In particular, the Purcell factor $P=\Gamma/\Gamma_0$, where $\Gamma_0$ is the decay rate in vacuum, is directly related to the ratio of the plasmon resonance quality factor to the mode volume. For a point dipole $\pb$ located at $\rb$, it can be calculated as \cite{Novotny}
\begin{equation}
P=1+\frac{3}{2p^2k_0^3}{\rm Im}\left\{\pb \cdot {\bf E}^{\rm ind}(\rb)\right\},
\nonumber
\end{equation}
where ${\bf E}^{\rm ind}(\rb)$ is the field induced by the dipole at its own position, which can be in turn obtained from the screened interaction as 
${\bf E}^{\rm ind}(\rb)=-\vec{\nabla}_\rb[(\pb \cdot \vec{\nabla}_{\rb'}){\WW}^{\rm ind}( \rb,\rb^{\prime},\omega)]\Big|_{\rb'=\rb}$. Using Eq.\ (\ref{W}) and specifying for dipoles oriented along the three orthogonal directions shown in Fig.\ \ref{Fig1}, we find
\begin{subequations}
\begin{equation}
P_\rho  =  1+\frac{3}{\pi k_0^3} \sum_{m=0}^{\infty}b_m \int_{0}^{\infty}\!\!\!\!dk \,k^2  \left[{K_m^{\prime}}(k\rho)\right]^2\;{\rm Im}\{r_m(k)\},
\end{equation}
\begin{equation}
P_z  =  1+\frac{3}{\pi k_0^3}\sum_{m=0}^{\infty}b_m \int_{0}^{\infty}\!\!\!\!dk\,k^2  K_m^2(k\rho)\;{\rm Im}\{r_m(k)\},
\end{equation}
\begin{equation}
P_\varphi  = 1+\frac{3}{\pi k_0^3} \sum_{m=1}^{\infty}c_m \int_{0}^{\infty}\!\!\!\!dk\,K_m^2(k\rho)\;{\rm Im}\{r_m(k)\},
\end{equation}
\label{PP}
\end{subequations}
\noindent where $K_m^{\prime}(z)=dK_m(z)/dz$, $b_m=(2-\delta_{m0})$, and $c_m=2m^2/\rho^2$.

\begin{figure}[htbp]
\includegraphics[width=\columnwidth]{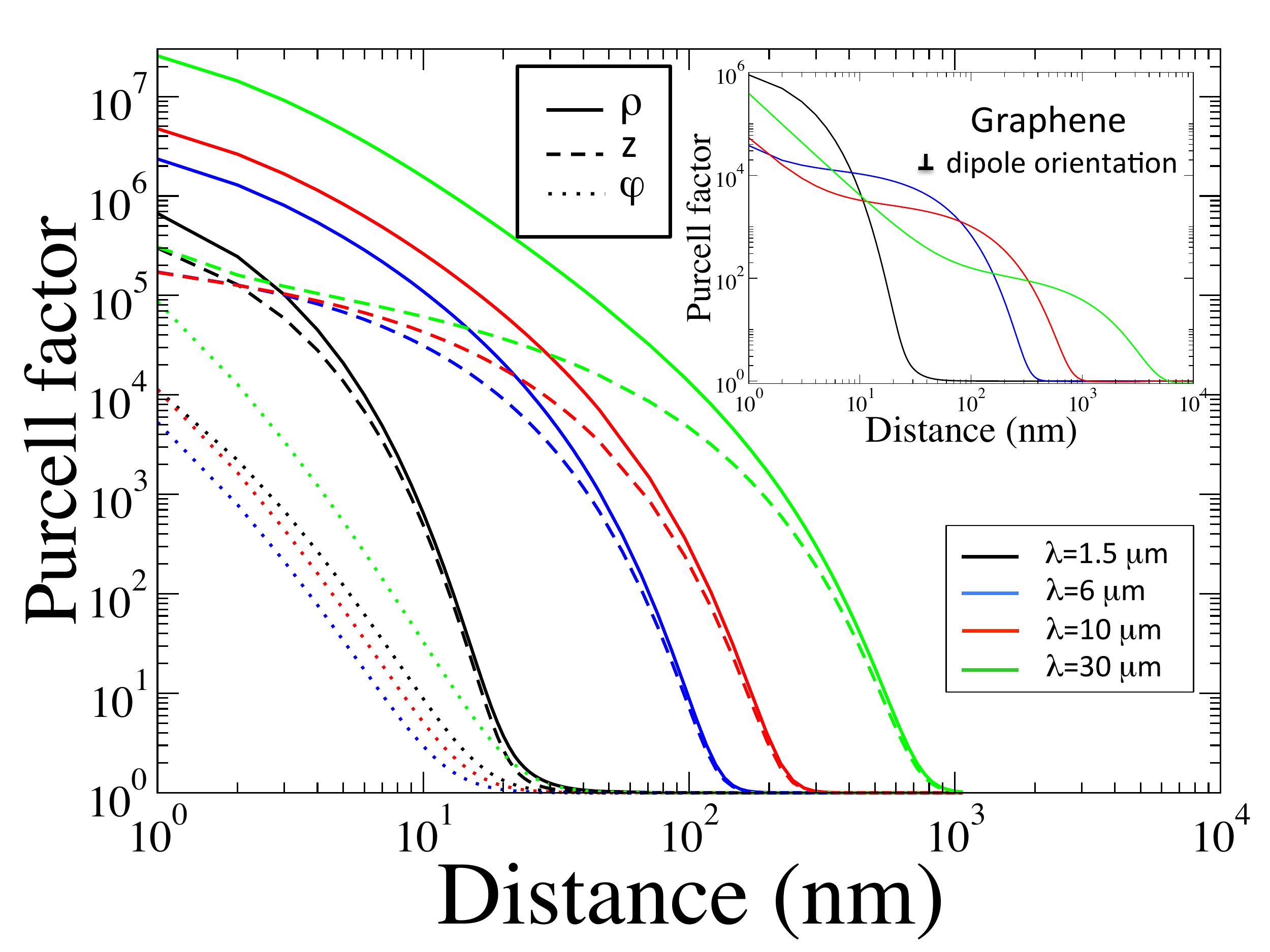}
\caption{{\bf Purcell factor.} The main panel shows the Purcell factor [Eqs.\ (\ref{PP})] as a function of the distance $\rho-R$ from the QE to the surface of a SWCNT of radius $R=2\,$nm for several values of the free-space emission wavelength $\lambda$ and all three possible QE dipole orientations (see Fig.\ \ref{Fig1}): radial (continuous curves), longitudinal (dashed curves), and azimuthal (dotted curves). The inset shows the corresponding Purcell factor in graphene for the same wavelengths and a dipole orientation perpendicular to the carbon plane.}
\label{Fig3}
\end{figure}

In Fig.\ \ref{Fig3}, we study the QE-SWCNT distance dependence of $P_\rho$, $P_\varphi$, and $P_z$ for tubes of radius $R=2\,$nm and for several QE emission wavelengths $\lambda$ in the $10-100\,\mu$m range. The highest Purcell factor is observed for radial orientation, as expected from the $-1/\rho$ divergence of the induced field at small separations, whereas the azimuthal orientation renders poor coupling because it is only contributed by $m\neq0$ SPs. When comparing the Purcell factor associated with either SWCNTs or graphene (see inset to Fig.\ \ref{Fig3}) for the corresponding optimal QE-dipole orientations ($P_\rho$ for SWCNTs and perpendicular to the 2D carbon sheet for graphene), it is clear that not only the Purcell factor is higher for 1D-SWCNTs than for 2D-graphene, but their spatial dependence is very different: whereas for graphene the Purcell factor increases rapidly towards short separations due to the dominant role of non-radiative channels, for SWCNTs this increase is much less pronounced.

\section{$\beta$ factor}

As mentioned above, a high coupling efficiency between QEs and propagating SPs is the key ingredient to achieve many of the proposed functionalities within the field of waveguide QED. In Figure 4 we show the $\beta$ factor as a function of the QE-SWCNT distance for several values of the QE-frequency. This magnitude can be evaluated by calculating the contribution of the SP-pole to the integrals of Eqs.\ (\ref{PP}). We evaluate the $\beta$ factor near a SWCNT of radius $R= 2$ nm for the two optimal QE-orientations, radial ($\rho$) and longitudinal ($z$). For comparison, we plot in the inset the evolution of the coupling efficiency for a QE near a graphene layer, with its dipole oriented perpendicularly to the 2D carbon plane. The coupling efficiency for SWCNTs reaches $100\%$ over a very broad range of distances: for the optimal orientation (radial), $\beta$ is close to $1$ for distances ranging from $1\,$nm to $1\mu$m (spanning three orders of magnitude). Importantly, this high coupling efficiency extends over a very broad range of frequencies and is even larger in the very low frequency regime. Our results imply that quenching phenomenon (i.e., when the QE emission is dominated by non-radiative decay channels) only appears at very short distances ($<1\,$nm) for QEs coupled to SWCNTs. As opposite to other metallic systems, SWCNTs present the crucial advantage of not requiring a spacer to avoid quenching of QEs at small separations. We attribute this negligible contribution of quenching modes to the very small quantity of material that binds the EM fields to the carbon nanotube. In contrast, in graphene a high coupling efficiency is only observed within a much narrower range of distances, and quenching shows up at distances below $\sim10\,$nm for the optimum wavelength, $\lambda=10 \mu$m (see inset of Fig. 4).

\begin{center}
\begin{figure}[htbp]
\includegraphics[width=\columnwidth]{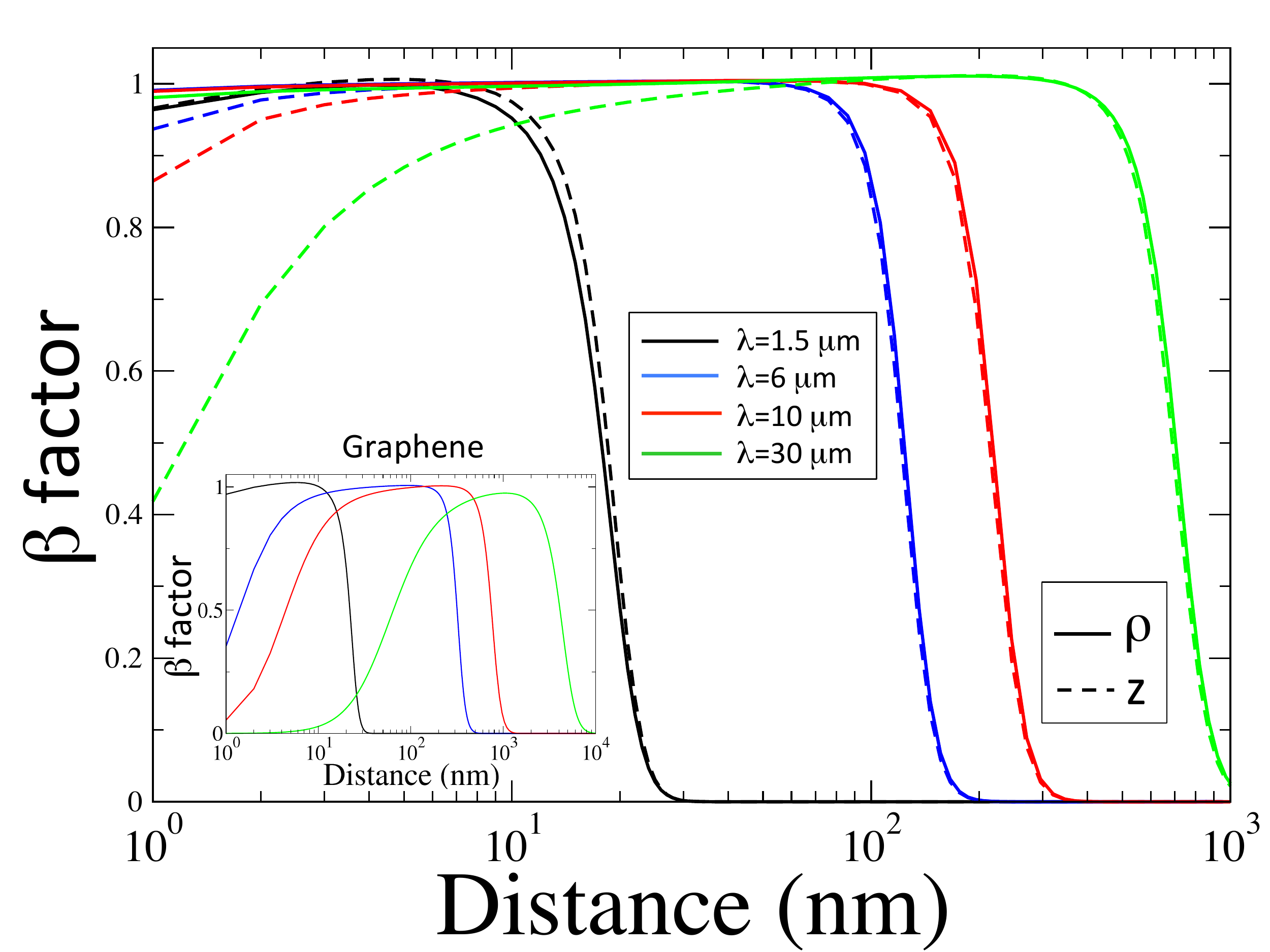}
\caption{{\bf $\beta$ factor.} Fraction of decay into plasmons ($\beta$ factor) from radially ($\parallel \rho$) and longitudinally ($\parallel z$) oriented QEs under the same conditions as in Fig.\ \ref{Fig3}. The inset shows the $\beta$ factor for a point dipole perpendicularly oriented to a 2D grapheme sheet.}
\label{Fig4}
\end{figure}
\end{center}

\section{Conclusion}

The strong interaction here predicted between QEs and the plasmons of SWCNTs opens new possibilities to implement waveguide QED schemes. It is worth noting that electrical contacts between gates and carbon nanotubes have been extensively studied for their potential as nanoelectronics elements, thus facilitating the design of practical electrical doping schemes. This combination of classical electrical tunability and efficient coupling with QEs constitutes a powerful platform for the investigation of fundamental quantum physics and the design of devices capable of processing information encoded in the states of the QEs.

\acknowledgments

This work is partially supported by the European Commission (Graphene Flagship CNECT-ICT-604391 and FP7-ICT-2013-613024-GRASP), by the European Research Council (ERC-2011-AdG Proposal No. 290981), and by the Spanish MINECO  under contract MAT2011-28581-C02.

\widetext

\appendix

\section{Quantum-mechanical versus classical calculations of the optical response}

Fig. S1 demonstrates that our classical level of description works reasonably well down to relatively small radius and short light wavelengths. In this figure we compare the classical description for the nanotube, in which its dielectric response is modeled by a local conductivity $\sigma(\omega)$ extracted from the local-RPA approximation (as in the main text), with a more microscopic theory. The latter is quantum-mechanical description that is based upon a tight-binding model for the electronic structure, used as input for the random-phase approximation \cite{Abajo14}. In particular, the graph is obtained using a previously developed numerical code \cite{Thon12} for plasmons in 1D graphitic structures with vanishing parallel wave vector. In both calculations the represented mode is the lowest-order one, which corresponds to transversal polarization of $m=1$ azimuthal symmetry. In the main paper, we find that the dominant contribution to the decay of QEs to be produced by longitudinal modes ($m=0$) integrated over a broad range of parallel wave vectors up to an effective cutoff $\sim1/\rho$, for which an even better level of accuracy is expected from the classical model, because the plasmon energies are then substantially below the Fermi energy, thus further away from the region in which coupling to electron-hole pairs can contribute to plasmon broadening and other nonlocal effects.

\begin{figure*}[b]
\centering
\includegraphics[width=10cm]{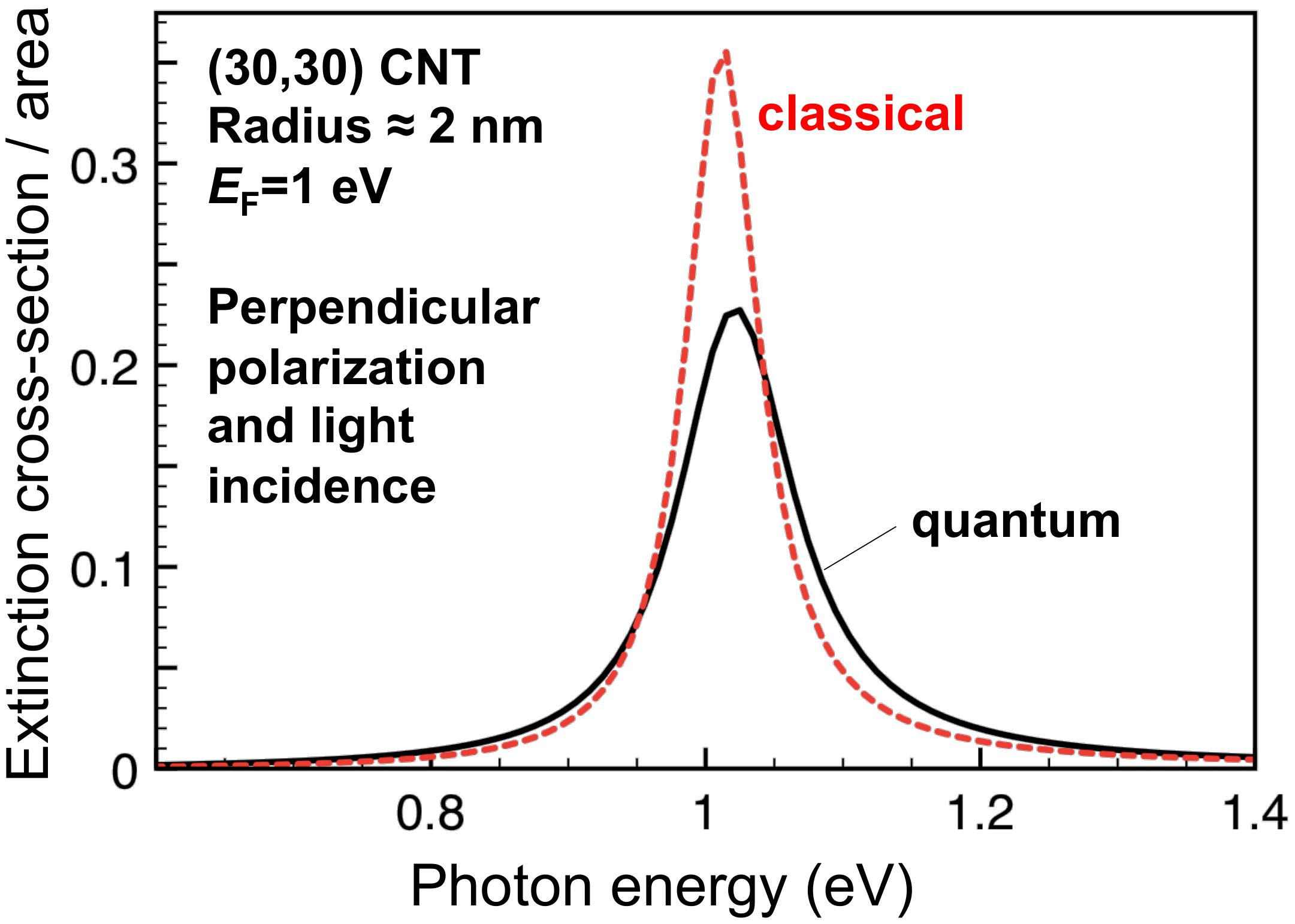}\\
\caption{{\bf Comparison between quantum-mechanical and classical descriptions of the optical response of highly-doped SWCNTs.} We represent the extinction cross-section normalized to the projected area of a $(30,30)$ SWCNT (radius $\approx2\,$nm) for light incidence and polarization directions both perpendicular to the tube axis. The Fermi energy is 1\,eV. The quantum-mechanical description produces a broad, slightly blue-shifted feature compared with the classical model, which is a characteristic nonlocal effect and disappears for plasmon modes of energy well below the Fermi energy.}
\end{figure*}

\section{Quasistatic approximation: Derivation of Eq.\ (1) of the main paper}

Electric and magnetic fields are decoupled in the quasi-static limit ($c\rightarrow\infty$), which is a reasonable approximation for structures such as the single-wall carbon nanotubes (SWCNTs) under consideration, where the dominant interaction between different regions of the material occurs at distances that are small compared with the light wavelength. In this limit, the magnetic field can be disregarded, while the electric field satisfies $\nabla\times\Eb=0$, so that it can be written as $\Eb=-\nabla\phi$ in terms of an electrostatic potential $\phi$. Using this expression, Coulomb's law reduces to the Laplace equation $\nabla^2\phi=0$ in the empty regions inside and outside the SWCNT. The response of the material then enters through the boundary conditions, namely, the continuity of the electric field components parallel to the tube, $\Eb_\parallel$, and the jump of $4\pi\nabla\cdot\jb/\ii\omega$ in the remaining perpendicular component right at the surface \cite{J99}, where $\jb=\sigma\cdot\Eb_\parallel$ is the induced current and $\sigma$ is the 2D conductivity.

We now present a succinct derivation of Eq.\ (1) of the main paper (the screened interaction $W^{\rm ind}$), from which the Purcell factor [Eq.\ (3)] readily follows from Eq.\ (2). We can calculate $W^{\rm ind}(\rb,\rb',\omega)$ as the potential induced at the position $\rb$ by a point charge placed at $\rb'$ and oscillating with frequency $\omega$. We implicitly assume a $\ee^{-\ii\omega t}$ time dependence in what follows and evaluate the conductivity $\sigma$ at that frequency. Given the symmetry of the structure, we use cylindrical coordinates $\rb=(\rho,z,\varphi)$, with the tube surface defined by $\rho=R$. As we study the interaction of the SWCNT with dipoles located outside it, we only work out the solution for $\rho,\rho'>R$.

We start by considering a complete basis set of solutions of the Laplace equation, $\ee^{\ii k z+\ii m\varphi}K_m(k\rho)$ and $\ee^{\ii kz+\ii m\varphi}I_m(k\rho)$, which are labeled by the real wave vector $k$ along the $z$ direction and the azimuthal number $m$. The bare Coulomb interaction then appears as a projection over those solutions,
\begin{eqnarray}
\frac{1}{|\rb-\rb'|}&=&\frac{1}{\pi}\int_{-\infty}^\infty dk\,\ee^{\ii k(z-z')}\sum_{m=-\infty}^\infty \ee^{\ii m(\varphi-\varphi')}
 \,\, I_m(k\rho_<)\,K_m(k\rho_>),
\label{eq:Coulomb}
\end{eqnarray}
where $\rho_<=\min\{\rho,\rho'\}$ and $\rho_>=\max\{\rho,\rho'\}$. When $\nabla^2$ is applied to the right-hand side of this equation, the jump at $\rho=\rho'$ produces a $\delta(\rho-\rho')$ function, which can be readily combined with the closure relations $\sum_m\ee^{\ii m(\varphi-\varphi')}=2\pi\delta(\varphi-\varphi')$ and $\int dk \ee^{\ii k(z-z')}=2\pi\delta(z-z')$ to verify that this is indeed the Green function of the Laplace equation [i.e., $\nabla^2(1/|\rb-\rb'|)=-4\pi\delta(\rb-\rb')$].

In order to calculate the induced interaction we need to consider $\rho_>=\rho'$ for the point charge and $\rho_<=\rho$ near the tube surface, so we study the response to each $\ee^{\ii k z+\ii m\varphi}I_m(k\rho)$ component, which we regard as an external potential. Because of the axial symmetry, the corresponding induced potential must conserve $k$ and $m$, and it must be $r_m\ee^{\ii kz+\ii m\varphi}K_m(k\rho)$ and $t_m\ee^{\ii kz+\ii m\varphi}I_m(k\rho)$ outside and inside the tube, respectively. Any other combination of basis-set solutions produces divergences at either $\rho=0$ or $\rho\rightarrow\infty$.  These expressions implicitly define reflection and transmission coefficients $r_m$ and $t_m$, which are determined by the boundary conditions at $\rho=R$. In particular, the continuity of $\Eb_\parallel$ also leads to a continuous potential, or equivalently, $t_mK_m(kR)=r_mI_m(kR)$, while the jump in the radial electric field gives $k r_mK'_m(kR)-k t_mI'_m(kR)=(4\pi\ii\sigma/\omega)(k^2+m^2/\rho^2)(t_m+1)I_m(kR)$, where the prime denotes differentiation with respect to the argument. The solution to these two linear equations produces
\begin{equation}
r_m=\frac{-I_m^2(kR)\Delta_m}{1+I_m(kR)K_m(kR)\Delta_m}, \nonumber
\end{equation}
where $\Delta_m=(4\pi\ii\sigma/\omega R)(m^2+k^2R^2)$. It should be noted that the derivation of this expression is facilitated by the use of the Wronskian $K_m(z)I'_m(z)-K'_m(z)I_m(z)=1/z$ \cite{AS1972}. Finally, the induced interaction is obtained from the right-hand side of Eq.\ (\ref{eq:Coulomb}) by replacing $r_mI_m$ for $I_m$, leading to
\begin{align}
W^{\rm ind}(\rb,\rb',\omega) \!=\! \frac{2}{\pi}\sum_{m=0}^{\infty}(2-\delta_{m0})\cos[m(\varphi-\varphi^{\prime})]
\int_{0}^{\infty}\!\!\!\!dk\;r_m(k)\;\cos[k(z-z^{\prime})]\; K_m(k\rho)K_m(k\rho'), \nonumber
\end{align}
[i.e., Eq.\ (1) of the main paper] after using the $k\rightarrow-k$ and $m\rightarrow-m$ symmetries of the integrand.
\begin{figure*}[htbp]
\centering
\includegraphics[width=10cm]{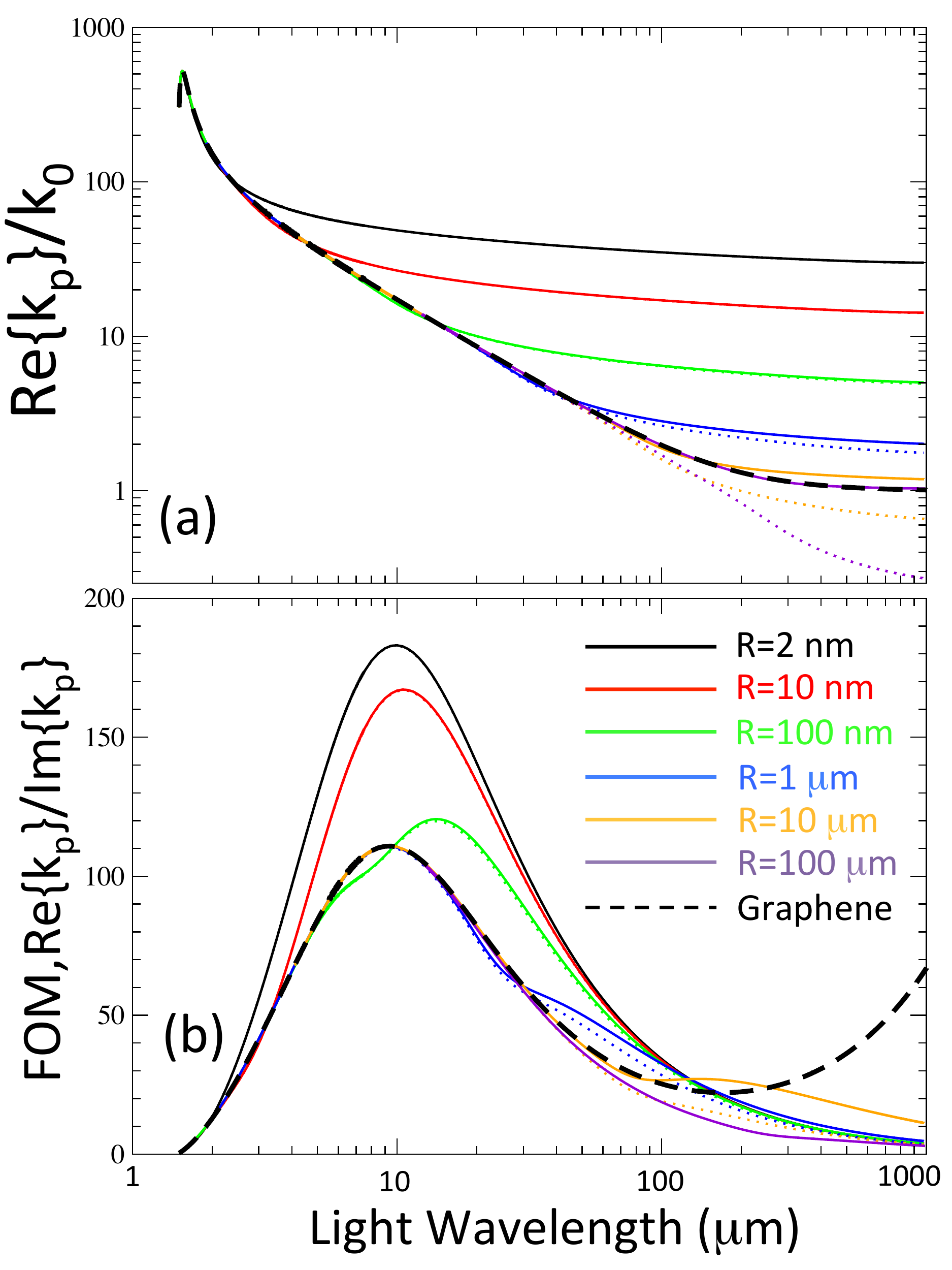}
\caption{{\bf Propagation characteristics of plasmons supported by SWCNTs and the limit towards graphene.} {\bf (a)} Real part of the plasmon wave vector $\kp$ as a function of the wavelength $\lambda$ of free-space light oscillating at the same frequency for different values of tube radius $R$. The plasmon wave vector is normalized to the light wave vector $k_0=2\pi/\lambda$. The limit of 2D graphene (dashed curve) is smoothly approach at large $R$'s. {\bf (b)} The corresponding figure of merit (FOM) ${\rm Re}\{\kp\}/{\rm Im}\{\kp\}$ of the guided plasmons studied in (a). Full electromagnetic theory (solid curves) is compared with the electrostatic limit (dotted curves) in both panels.}
\end{figure*}

\begin{figure*}[htbp]
\centering
\includegraphics[width=10cm]{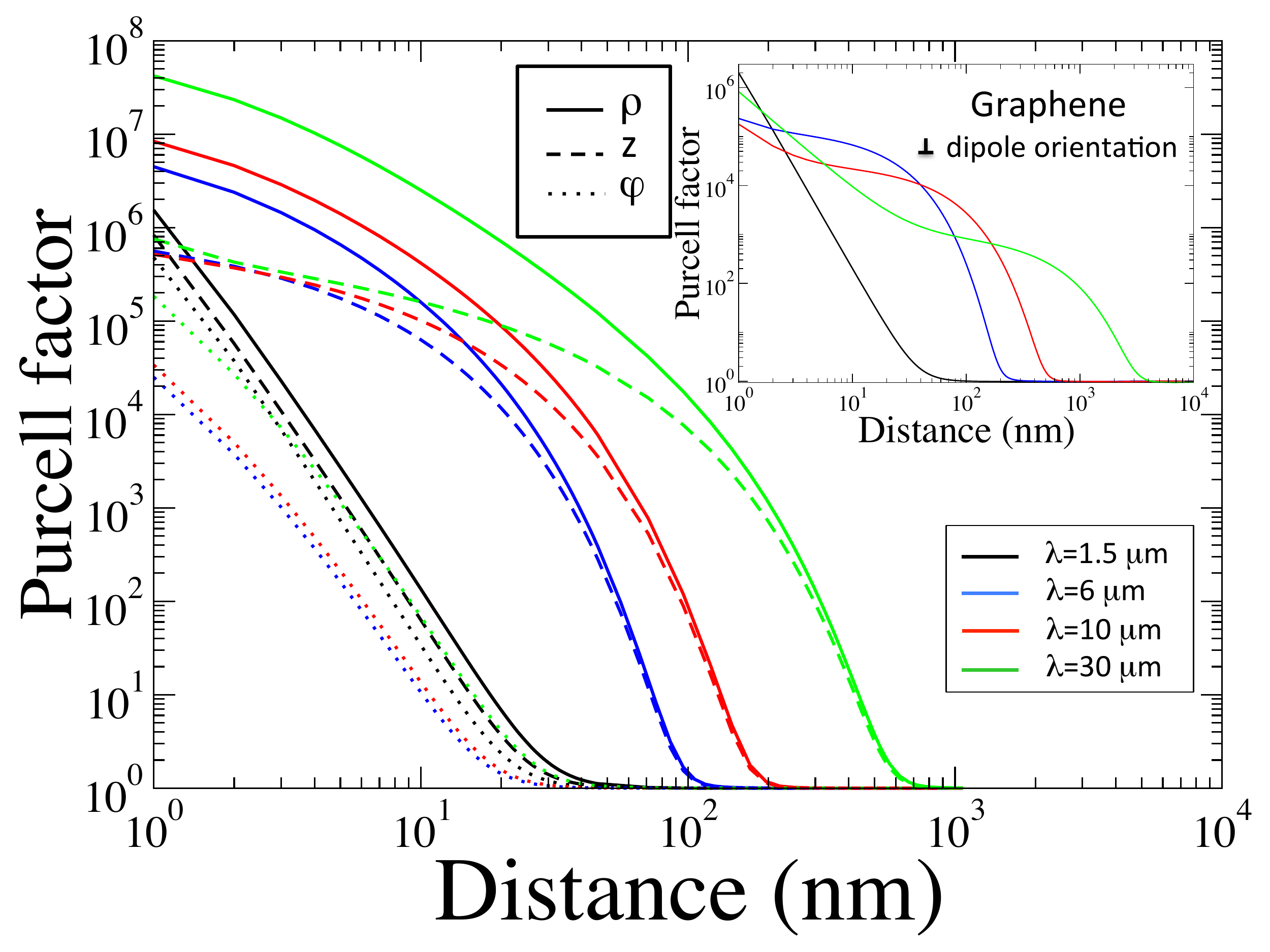}
\caption{{\bf Purcell factor.} The main panel shows the Purcell factor as a function of the distance $\rho-R$ from the QE to the surface of a SWCNT of radius $R=2\,$nm for several values of the free-space emission wavelength $\lambda$ and all three possible QE dipole orientations (see Figure 1 in the main text): radial (continuous curves), longitudinal (dashed curves), and azimuthal (dotted curves). The inset shows the corresponding Purcell factor in graphene for the same wavelengths and a dipole orientation perpendicular to the carbon plane.}
\end{figure*}

\begin{figure*}[htbp]
\centering
\includegraphics[width=10cm]{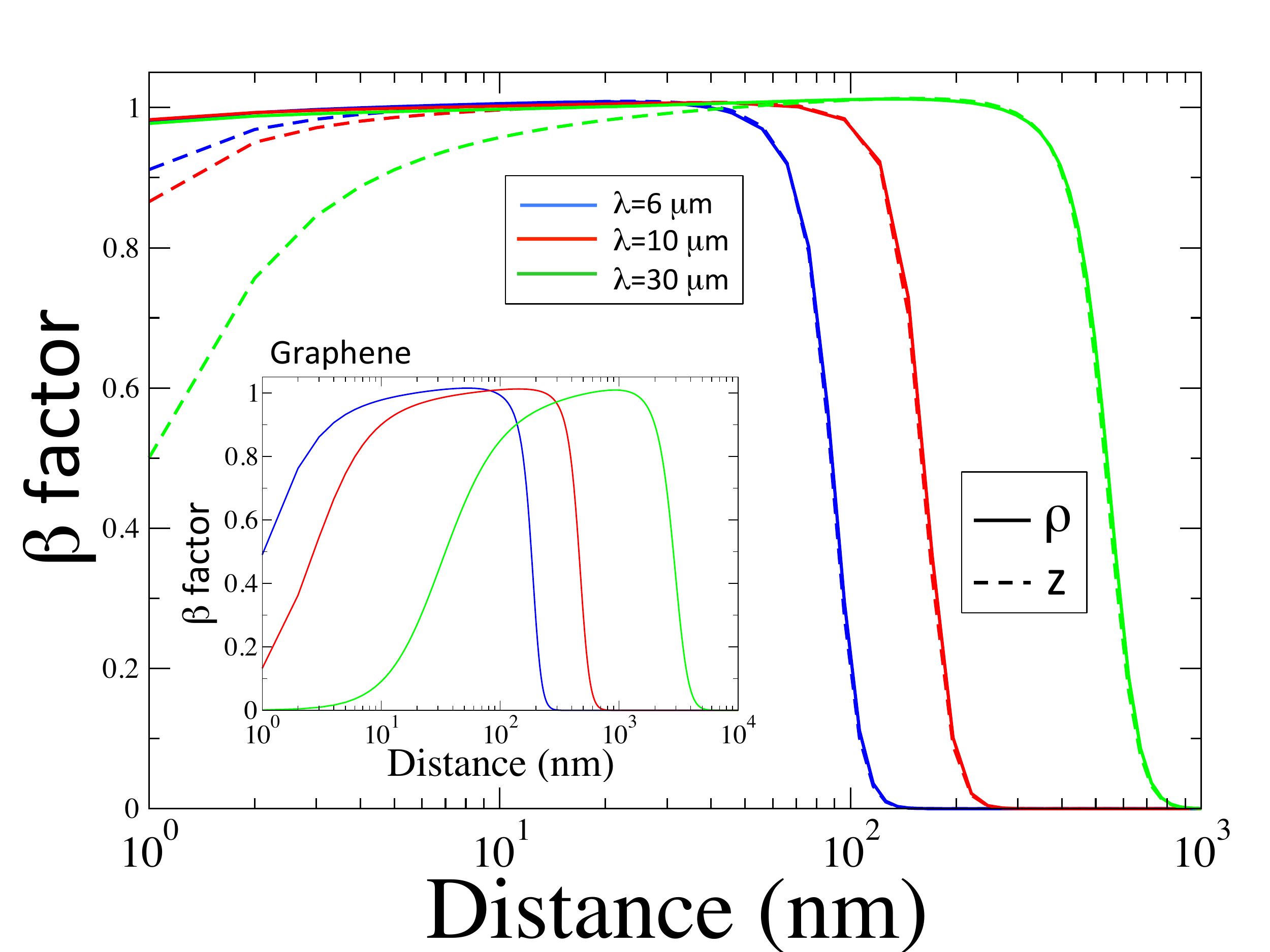}
\caption{{\bf $\beta$ factor.} Fraction of decay into plasmons ($\beta$ factor) from radially ($\parallel \rho$) and longitudinally ($\parallel z$) oriented QEs under the same conditions as in Fig.S3. The inset shows the $\beta$ factor for a point dipole perpendicularly oriented to a 2D graphene sheet.}
\end{figure*}

\section{Results for a low doping level}

In this section we present similar results to those of the main text, but for a different doping level in the SWCNT. In Figures S2, S3 and S4, which are the counterparts for Figures 2, 3 and 4 of the main text, we use $E_F=0.5$ eV instead, whereas the other parameters are exactly the same as those utilized in the figures of the main text. A conclusion of this analysis is that the capabilities of SWCNTs for ultra-efficient coupling to quantum emitters are very robust against changes in the doping level. Notice that in Fig. S4 we omit the calculation for $\lambda=1.5 \mu$m, as there are no propagating surface plasmons supported by either graphene or SWCNTs for $E_F=0.5$ eV at that wavelength (see Fig. S2).  

\clearpage

\section{Full electrodynamic calculation}

\subsection{M and N functions}

The calculation of the electromagnetic properties in a system with cylindrical symmetry is facilitated by the introduction of two base functions:
\begin{eqnarray}
\M(\vec{r})& = &\nabla \times \left[f(\vec{r}) \, \vec{u}_z \right],    \\ \nonumber
\N(\vec{r}) & = & \frac{1}{g_k} \, \nabla \times \M(\vec{r}),
\end{eqnarray}
where  $f(\vec{r}) = J_n(k r) \, e^{\imath (n \theta + k_z z)}$, $n$ is an integer, and $g_k = \sqrt{k^2+k_z^2}.$ \\

The $N$ function corresponds to the TM polarization, while the $M$ function is a TE mode. These functions satisfy
\begin{eqnarray}
\nabla \times \M &=& g_k \, \N \\ \nonumber
\nabla \times \N &=& g_k \, \M
\end{eqnarray} \\
and, in the coordinate system $\left( \vec{u}_{r}, \vec{u}_{\theta}, \vec{u}_{z} \right)$, they admit the following explicit expressions:
 
 \begin{equation}
\M(\vec{r}) = \begin{pmatrix}
 \dfrac{\imath n}{r} \, J_{n}(kr)  \\[0.6 em] 
- k  \, J'_{n}(kr)   \\[0.6 em] 
0
 \end{pmatrix}  \,  e^{\imath n \theta} \, e^{\imath k_z z}, 
 \hspace{1cm}
 \N(\vec{r}) = \dfrac{1}{g_k} \, \begin{pmatrix}
\imath k k_{z} \, J'_{n}(kr)   \\[0.6 em]  
 - \dfrac{n k_{z}}{r} \, J_{n}(kr)  \\[0.7 em] 
 k^{2}  \, J_{n}(kr) 
 \end{pmatrix}  \,  e^{\imath n \theta} \, e^{\imath k_z z}.
 \end{equation} \\

These wave functions are related to the even-odd functions found in the literature (for instance in Ref. \cite{Chen}): 
$\vec{M} = \vec{M}_e + \imath \vec{M}_o$, $\vec{N} = \vec{N}_e + \imath \vec{N}_o$. However the even-odd functions contain different angular dependences for different field components, and furthermore, they are plagued by changes of sign between even and odd components. The single-exponential functions are simpler, but contain the subtlety that simple conjugation does not provide an orthogonal set of functions. 

To perform the calculations we need {\it orthogonal} wave functions, which we find by inspection to be:
 \begin{eqnarray}
\Md(\vec{r}) &=& \left(
- \dfrac{\imath n}{r} \, J_{n}(kr) , - k  \, J'_{n}(kr)  , 0 \right) \,  e^{-\imath n \theta} \, e^{-\imath k_z z} \\ \nonumber 
 \Nd(\vec{r}) &=& \dfrac{1}{g_k} \, \left(- \imath k k_{z} \, J'_{n}(kr),  - \dfrac{n k_{z}}{r} \, J_{n}(kr), 
 k^{2}  \, J_{n}(kr) 
 \right)  \,  e^{-\imath n \theta} \, e^{-\imath k_z z}
 \end{eqnarray} 

with the orthogonality relations:
 \begin{eqnarray}
\int d\vec{r} \, \Mdp(\vec{r})  \cdot \M(\vec{r}) = \int d\vec{r} \, \Ndp(\vec{r}) \cdot \N(\vec{r}) &=&
4 \pi^2 k \, \delta(k-k')\, \delta(k_z-k'_z) \, \delta_{n,n'}, \\ \nonumber
\int d\vec{r} \, \Ndp(\vec{r}) \cdot \M(\vec{r})= \int d\vec{r} \, \Ndp(\vec{r}) \cdot \M(\vec{r}) &=&0.
 \end{eqnarray} 

At this point we find useful to simplify the notation: we define $\ket{\Psi(o)}$, 
where $o \equiv \left\{ \sigma, n, k, k_z\right\}$,
and $\bar{o} \equiv \left\{ \bar{\sigma}, n, k, k_z\right\}$, and $\sigma = M, N$, such that
\begin{equation}
\braket{\vec{r}|\Psi_{o}} = \M(\vec{r}), \hspace{1cm} \braket{\Psi_{o}| \vec{r}} = \Md(\vec{r})
\end{equation}
for $\sigma=1=M$, and a similar equation interchanging $M$ and $N$ for $\sigma=2=N$.

The operators may act over different labels of $\Psi$. For instance, the relations between $M$ and $N$ and the orthonormality expressions are written as
\begin{eqnarray}
\braket{\Psi_{o}| \Psi_{o'}} &=& 4 \pi^2 k \, \delta(o-o') \equiv N_o \, \delta(o-o'), \nonumber \\
\nabla \times \ket{\Psi_{o}} &=& g_k \ket{\Psi_{\bar{o}}},  
\label{eq:orthog}  
 \end{eqnarray} 
 where $N_o$ is an overlap factor (which could be included in the normalization of the functions $M$ and $N$).
 
We will also use the compact notation
\begin{equation}
\int do \equiv \sum_{\sigma=M,N} \sum_{n=-\infty}^{\infty} \int_0^{\infty}  dk \int_{-\infty}^{\infty}  dk_z.
\end{equation}

\subsection{Vacuum Green's Function}

There are two dyadic Green's functions commonly used, the magnetic and the electric ones, which are defined by:
 \begin{eqnarray}
 \nabla \times \nabla \times \Ge - g^2 \, \Ge &=& \mathbb{1}, \\ \nonumber
 \nabla \times \nabla \times \Gm - g^2 \, \Gm &=& \nabla \times \mathbb{1}, 
 \label{eq:GeGm}
 \end{eqnarray} 
 where $g = \omega/c$ is the free space wavevector. 
 
It is more convenient to start by computing ${\overline{\overline{G}}_m}$, and derive ${\overline{\overline{G}}_e}$ from it (Eq. 5.156 in \cite{Chen}):
\begin{equation}
\Ge=\dfrac{1}{g^{2}} \, \left( \nabla \times \Gm \, - \vec{u}_{r} \vec{u}_{r}. \right)
\end{equation} 

To compute $\Gm$ we represent it on the chosen basis set: 
\begin{equation}
\Gm = \int \int \, do \, do' \, a_{o,o'} \ket{\Psi_o}\bra{\Psi_{o'}}.
\end{equation}
Using   Eq.(S8) and Eq.(S10), 
we find  $ \left( g_k^2 - g^2 \right) \Gm = \nabla \times \mathbb{1} $, and projecting on
$\bra{\Psi_x}$ from the left and $\ket{\Psi_y}$ from the right, we obtain:
\begin{equation}
N_{x} N_{y} \left( g_{k_{x}}^2 - g^2  \right) a_{x,y} =  
g_{k_{y}} \braket{\Psi_{x}|\Psi_{\bar{y}}} =  g_{k_{y}} N_{y} \delta_{x,\bar{y}}
\nonumber 
\end{equation}
Therefore
\begin{equation}
a_{o,o'} = \frac{g_{o}}{g_{o}^2 - g^2} \, \frac{1}{N_o} \, \delta_{o,o'}
\end{equation}
and
\begin{equation}
\Gm = \int \, do \, \frac{1}{N_o} \, \frac{g_{o}}{g_{o}^2 - g^2}  \, \ket{\Psi_{o}}\bra{\Psi_{o}}, \\
\end{equation}
Using $\Ge = -( \vec{u}_r  \vec{u}_r   + \nabla \times \Gm) / g^2$ and $N_o = N_{\bar{o}}$, $g_s = g_{\bar{o}}$, the expression for the electric dyadic Green's function is:
\begin{equation}
\Ge =   -\frac{\vec{u}_{r}  \vec{u}_{r}}{g^{2}} +  \int \, do \, \frac{1}{N_{o}} \, \frac{g_{o}^2}{g^2 \, (g_{o}^2 - g^2)}  \, \ket{\Psi_{o}} \bra{\Psi_{o}}.
\end{equation}

An important point is that the integral is over waves with all values for $k$ and $k_z$, even for those such that
$k^2+k_z^2 \equiv g_k^2 \neq g^2$, which are not 'in-shell'. Now, the integral over $k$ can be evaluated by contour integration in the complex $k$-plane, which leads to:
\begin{equation}
\int f(k) \frac{J_n(kr) J_n(k r')}{k^2 - \left( g^2-k_z^2 \right)} dk = \frac{\imath \pi f(k)}{2 k} 
\begin{cases} 
J_n(k r) H^{(1)} (k r'), & \mbox{if } r'>r, \\
H^{(1)} (k r') J_n(k r), & \mbox{if } r'<r,
\end{cases}
\end{equation}
where now $k$ is 'in-shell', i.e., it satisfies $k^2+k_z^2 = g^2$. \\

We now define an ``in-shell'' label $s \equiv \left\{ \sigma, n, k_z\right\}$, with $\int ds \equiv \sum_{\sigma=M,N} \sum_{n=-\infty}^{\infty} \int_{-\infty}^{\infty}  dk_z$. With this, we find
\begin{eqnarray}
\Gm &= &\frac{\imath g}{8\pi} \int \, ds \, \frac{1}{k^2}  \, \ket{\Psi_{s}}\bra{\Psi_{\bar{s}}}, \\ \nonumber
\Ge &= &  -\frac{1}{g^2} \vec{u}_r  \vec{u}_r \, + \,  \frac{\imath}{8\pi} \int \, ds \, \frac{1}{k^2}  \, \ket{\Psi_s}\bra{\Psi_s}.
\end{eqnarray} 

Recall that these expressions come from an integral over $k$, performed in the complex plane. 
When closing the contour, different contributions $J, H^{(1)}$, etc. appear depending on the relation between $r$ and $r'$. Therefore, the expression
$\braket{\vec{r}|\Ge|\vec{r'}}$ must be understood as: 
\begin{equation}
\braket{\vec{r}|\Ge|\vec{r'}} \equiv  \braket{\vec{r}|\Psi_s} \braket{\Psi_s|\vec{r'}} \, =  
\begin{cases} 
\Psi_s(\vec{r}) \, \Psi_s^{(1)} (\vec{r'}), & \mbox{if } r'>r, \\
\Psi_s^{(1)} (\vec{r'}) \, \Psi_s(\vec{r}),  & \mbox{if } r'<r,
\end{cases}
\end{equation}
where $ \Psi_s^{(1)} $ is like $\Psi_s(\vec{r})$ by replacing $J \rightarrow H^{(1)}$.

We are interested on the dyadic Green's functions for sources outside the nanotube, 
therefore the relevant vacuum Green's functions are:
\begin{eqnarray}
\vGe &=&  -\frac{1}{g^2} \vec{u}_r  \vec{u}_r \, + \,  \frac{\imath}{8\pi} \int ds \, \frac{1}{k^2}  \, \ket{\Psi_s}\bra{\Psi_s^{(1)}}, \\ \nonumber 
\vGm &=&   \frac{\imath g}{8 \pi} \int ds \, \frac{1}{k^2}  \, \ket{\Psi_{\bar{s}}}
\bra{\Psi_s^{(1)}}.
\label{eq:VGD}
\end{eqnarray} 

\subsection{Contribution to the Green's function from reflected waves}

On top of the vacuum dyadic Green's function, we must add a contribution in order to satisfy the boundary conditions at the SWCNT. In the region outside the SWCNT these contributions are associated to the reflexion of the "incident" wave $\ket{\Psi_s}$, which can be written as:
\begin{eqnarray}
\rGe & = &   \frac{\imath}{8\pi}  \int \int \, ds \, ds' \, \frac{1}{k^2}  \, \ket{\Psi_{s}^{(1)}} r_{s,s'} \bra{\Psi_{s'}^{(1)}},
 \\ \nonumber
 \rGm &= & \frac{\imath g }{8\pi} \int \int \, ds \, ds' \, \frac{1}{k^2}  \, \ket{\Psi_{\bar{s}}^{(1)}} r_{s,s'}  \bra{\Psi_{s'}^{(1)}}
 \label{eq:RGD}
\end{eqnarray} 
with some coefficients $r_{s,s'}$ that must be determined form SWCNT properties.

In the case of a SWCNT, rotational and translational symmetries imply
\begin{equation}
r_{s,s'} = r_{\sigma, \sigma'} \, \delta_{n,n'} \, \delta_{k_{z}, k_{z}'}
\end{equation}
and therefore
\begin{eqnarray}
\rGe & = &   \frac{\imath}{8\pi} \sum_{n=-\infty}^{\infty} \,  \sum_{\sigma, \sigma'=M,N} 
\int_{-\infty}^{\infty} dk_{z} \, \frac{1}{k^2}  \, r_{\sigma,\sigma'} \, \ket{\Psi_{\sigma,n,k_{z}}^{(1)}}\bra{\Psi_{\sigma',n,k_{z}}^{(1)}},
 \\ \nonumber
 \rGm &= & \frac{\imath g }{8\pi} \sum_{n=-\infty}^{\infty} \,  \sum_{\sigma, \sigma'=M,N} 
 \int_{-\infty}^{\infty} dk_{z} \, \frac{1}{k^2}  \, r_{\sigma,\sigma'} \, \ket{\Psi_{\bar{\sigma},n,k_{z}}^{(1)}}\bra{\Psi_{\sigma',n,k_{z}}^{(1)}}.
 \label{eq:RGD}
\end{eqnarray} 

Inside the SWCNT, the fields should go as $\ket{\Psi_{\bar{\sigma},n,k_{z}} }$ (they do not have to satisfy the radiation condition at $r \rightarrow \infty$). 
Taking into account rotational symmetry the expression for the dyadic Green's function can be expressed in terms of the ``transmission'' amplitudes $ t_{\sigma,\sigma'}$ as:
\begin{eqnarray}
\tGe & = &   \frac{\imath}{8\pi} \sum_{n=-\infty}^{\infty} \,  \sum_{\sigma, \sigma'=M,N} 
\int_{-\infty}^{\infty} dk_{z} \, \frac{1}{k^2}  \, t_{\sigma,\sigma'} \, \ket{\Psi_{\sigma,n,k_{z}}}
\bra{\Psi_{\sigma',n,k_{z}}^{(1)}},
 \\ \nonumber
 \tGm &= & \frac{\imath g }{8\pi} \sum_{n=-\infty}^{\infty} \,  \sum_{\sigma, \sigma'=M,N} 
 \int_{-\infty}^{\infty}dk_{z} \, \frac{1}{k^2}  \, t_{\sigma,\sigma'} \, \ket{\Psi_{\bar{\sigma},n,k_{z}}}\bra{\Psi_{\sigma',n,k_{z}}^{(1)}}.
 \label{eq:TGD}
\end{eqnarray} 

In order to find the coefficients $r_{\sigma,\sigma'}, t_{\sigma,\sigma'}$ we must impose the boundary conditions at $r=R$. 
\begin{eqnarray}
\vec{u}_{r} \times \left(\vec{E}_{1}-\vec{E}_{2}\right) & = & 0 \\ \nonumber
\vec{u}_{r} \times \left(\vec{H}_{1}-\vec{H}_{2}\right) & = & 2 \alpha \vec{E}
\end{eqnarray}

\noindent where $\alpha$ is link to the local conductivity of the SWCNT, $\sigma$, $\alpha=\frac{2\pi \sigma}{c}$. 

The magnetic field is related to the electric field through 
$\vec{H} = -(\imath/g) \nabla \times \vec{E}$. On the other hand, from Eq.\ref{eq:VGD} and \ref{eq:RGD} we know that $\Gm = \nabla \times \Ge$. Thus, the magnetic field generated by a point {\it electric} dipole is:
\begin{equation}
\vec{H}  =  -\frac{\imath}{g} \, \Gm
\end{equation}
and, thus the boundary condition in a SWCNT at the location of the carbon sheet is
\begin{eqnarray}
\vec{u}_{r} \times \left(\Geone-\Getwo \right) &=& 0 \\ \nonumber
\vec{u}_{r} \times \frac{1}{g} \left(\Gmone-\Gmtwo\right) &=& 2 \imath \alpha \, \Ge.
\end{eqnarray}

\subsubsection{Reflection coefficients}

\begin{itemize}
\item Incidence by a wave $\ket{M}$
\begin{itemize} 
\item $\vec{u}_{r} \times \left(\Geone-\Getwo \right)=0$
\begin{eqnarray}
 M_{2} + r_{MM} M_{2}^{(1)} + r_{NM} N_{2}^{(1)} &=& t_{NM} N_{2} + t_{MM} M_{2} \\ \nonumber
  M_{3} + r_{MM} M_{3}^{(1)} + r_{NM} N_{3}^{(1)} &=& t_{NM} N_{3} + t_{MM} M_{3} 
\end{eqnarray}
 
 \item $\vec{u}_{r} \times (\Gmone-\Gmtwo) / g = \alphabar \Gm$
 \begin{eqnarray}
- \left[ N_{3} + r_{MM} N_{3}^{(1)} + r_{NM} M_{3}^{(1)}- t_{NM} M_{3} - t_{MM} N_{3} \right]  &=&  \alphabar ( t_{NM} N_{2} + t_{MM} M_{2}) \\ \nonumber
  N_{2} + r_{MM} N_{2}^{(1)} + r_{NM} M_{2}^{(1)}- t_{NM} M_{2} - t_{MM} N_{2}  &=&  \alphabar ( t_{NM} N_{3} + t_{MM} M_{3}), 
\end{eqnarray}
\end{itemize}
where all components of $N$ and $M$ are to be evaluated at $r=R$, 
and $\alphabar \equiv 2 \imath \alpha$. 

In matrix form \\

$  \underbrace{\begin{pmatrix} 
 M_{2}^{(1)} & N_{2}^{(1)} & - M_{2} & -N_{2} \\
 M_{3}^{(1)} & N_{3}^{(1)} & - M_{3} & -N_{3} \\
 N_{3}^{(1)} & M_{3}^{(1)} & - N_{3} + \alphabar M_{2} & -M_{3} + \alphabar N_{2} \\
 N_{2}^{(1)} & M_{2}^{(1)} & - N_{2} - \alphabar M_{3} & -M_{2} - \alphabar N_{3} \\
 \end{pmatrix}  }_{A}
 \underbrace{ \begin{pmatrix}  r_{MM} \\ r_{NM} \\ t_{MM} \\ t_{NM} \end{pmatrix}}_{C}   =
 \underbrace{\begin{pmatrix}  -M_{2} \\ -M_{3}  \\ N_{3}+\alphabar M_{2}\\ -N_{2}+\alphabar M_{3}  \end{pmatrix}}_{B} $ \\

\item Incidence by a wave $\ket{N}$

For this incidence a similar set of equations is obtained, with the same square matrix $A$, the vector of coefficients $C= \begin{pmatrix}  r_{MN}, r_{NN} , t_{MN} , t_{NN} \end{pmatrix}^{T}$,
and a vector $B$ with the same expression as before, but with the substitution $ N \longleftrightarrow M$. 
\end{itemize}

These systems of equations can be solved analytically. Defining adimensional quantities 
$\tilde{R} \equiv g R$, $q\equiv k/g$ and $q_z\equiv  k_z/g$, 
we obtain:
\begin{eqnarray}
D &=& \alphai \bhn \bjn (\bhpn \bjn - \bhn \bjpn) n^2 q_{z}^2 + \\ \nonumber
    & & q^2 (\bhpn \bjn - \bhn \bjpn + \alphai \bhn \bjn q) (\alphai \bhpn \bjpn + q(
     \bhpn \bjn - \bhn \bjpn)) \tilde{R}^2 \\ \nonumber \\ \nonumber
r_{MM} &=&  -\frac{\alphai \bjpn^2 q^2 (\bhpn \bjn - \bhn \bjpn + \alphai \bhn \bjn q) \tilde{R}^2}{D} \\ \nonumber
r_{NM} &=&  r_{MN} =    \frac{\alphai \bjn \bjpn (-\bhpn \bjn + \bhn \bjpn) n q q_{z} \tilde{R}}{D} \\ \nonumber
r_{NN} &=&\frac{\alphai \bjn^2 ((-\bhpn \bjn + \bhn \bjpn) n^2 q_{z}^2 - 
   q^3 (\alphai \bhpn \bjpn + \bhpn \bjn q - \bhn \bjpn q) \tilde{R}^2))}{D}     
\end{eqnarray}

which, using the Wronskian relation:
\begin{equation}
J_{n} (H_{n}^{(1)})'(x)- J'_{n}(x) H_{n}^{(1)}(x) = \frac{2 \imath }{\pi x} 
\end{equation}

simplifies to
\begin{eqnarray}
D &=& \alphai \bhn \bjn  W n^2 q_{z}^2 + q^2 (W + \alphai \bhn \bjn q) (\alphai \bhpn \bjpn + q W) \tilde{R}^2 \\ \nonumber \\ \nonumber 
r_{MM} &=&  -\frac{\alphai {\bjpn}^2 q^2 (W + \alphai \bhn \bjn q) \tilde{R}^2}{D} \\ \nonumber
r_{NM} &=&  r_{MN} =  - \frac{n \alphai \tilde{R} q q_{z}  W \bjn \bjpn }{D} \\ \nonumber
r_{NN} &=&  - \frac{\alphai {\bjn}^2 (n^2 q_{z}^2 W + q^3 \tilde{R}^2 (\alphai \bhpn \bjpn +q W)}{D}, 
\end{eqnarray}
where $W = \frac{2 \imath }{\pi q \tilde{R}}$. Multiplying all numerators and denominator by $W^{-2}$, we obtain:\\
\begin{subequations}
\begin{empheq}[box=\widefbox]{align}
D &= n^2 \pi \alpha \, q_{z}^2 q \tilde{R} \bjn \bhn + q^3 \tilde{R}^2 (1 + \pi \alpha \, q^2 \tilde{R} \bjn \bhn ) (1+ \pi \alpha \, \tilde{R} \bjpn \bhpn)  \\ \nonumber \\ \nonumber 
r_{MM} &=  -\frac{\pi \alpha \, q^3  \tilde{R}^3  \bjpn^2 (1+ \pi \alpha \, q^2 \tilde{R}  \bjn  \bhn)}{D} \\ \nonumber
r_{NM} &=   r_{MN} = - \frac{n \pi \alpha \, q_{z}  q^2  \tilde{R}^2 \bjn \bjpn }{D} \\ \nonumber
r_{NN} &= - \frac{\pi \alpha \, q \tilde{R}{\bjn}^2 \,  (n^2 q_{z}^2 + q^4 \tilde{R}^2 (1+ \pi \alpha \, \tilde{R} \bjpn \bhpn))}{D}, 
\end{empheq}
\label{E:reflection-coeff}
\end{subequations}
where all Bessel functions have argument $q\tilde{R}$.

\subsection{Purcell factor}

In the CGS system, the spontaneous emission rate is 
\begin{equation}
\Gamma = \frac{1}{2g} \,  \mathrm{Im} \left\{ {\vec{p}}^\intercal  \cdot \Ge \cdot \vec{p} \right\}
\end{equation}

A simple calculation shows that in vacuum 
\begin{equation}
\Gamma_{vac} = \frac{1}{12 \pi}
\end{equation}

Therefore, the Purcell factor, defined as $P = \Gamma/\Gamma_{vac}$ can be computed from the reflected part of the dyadic Green's function as:
\begin{equation}
P = 1 + \frac{3}{4 g} \, \mathrm{Re} \left\{ \sum_{n=-\infty}^{\infty} \,  \sum_{\sigma, \sigma'=M,N} \int_{-\infty}^{\infty} dk_{z} \, \frac{1}{k^2}  \, r_{\sigma,\sigma'} \, \braket{\vec{p} | \Psi_{\sigma,n,k_{z}}^{(1)}} \braket{\Psi_{\sigma',n,k_{z}}^{(1)} | \vec{p}} \right\}
\end{equation}

As both $r_{N,N}$ and $r_{M,M}$ are even in $n$, but so are 
$ \braket{\vec{p} | \Psi_{M}} \braket{\Psi_{M} | \vec{p}}$ and   
$ \braket{\vec{p} | \Psi_{N}} \braket{\Psi_{N} | \vec{p}}$. 
The diagonal elements in $\sum_{\sigma, \sigma'=M,N}$ are even in $n$.  
The non-diagonal elements $r_{N,M}$ and $r_{M,N}$ are odd in $n$, but 
the components of   $ \braket{\vec{p} | \Psi_{M}}\braket{\Psi_{N} | \vec{p}}$ 
are also odd in $n$, so the partial contributions to the Purcell factor are all 
even in $n$. Similarly, using that, for all Bessel functions $J_{n}(-x) = -J_{n}(x)$ 
and the fact that the Bessel functions always appear in pairs, the integral 
over $k_{z}$ can be restricted to the interval $(0, \infty)$.

So finally, using adimensional units, we find

\begin{equation}
\boxed{P = 1 + \frac{3}{2} \, \mathrm{Re} \left\{ \sum_{n=0}^{\infty} \, (2-\delta_{n,0})  \sum_{\sigma, \sigma'=M,N} \int_{0}^{\infty} dq_{z} \, \frac{1}{q^2}  \, 
\left(\vec{p} \cdot  \vec{v}_{\sigma,n,q_{z}} \right) \,  r_{\sigma,\sigma'} \, 
\left( \vec{v}_{\sigma',n,q_{z}} \cdot \vec{p} \right) \right\}},
\end{equation}
where
\begin{equation}
\vec{v}_{M,n,q_{z}}= \begin{pmatrix}
 \dfrac{\imath n}{\tilde{d}} \, H_{n}^{(1)}(q \tilde{d})  \\[0.6 em] 
- q  \, H_{n}^{(1)'}(q \tilde{d})  \\[0.6 em] 
0
 \end{pmatrix}, 
 \hspace{1cm}
\vec{v}_{N,n,q_{z}}=  \begin{pmatrix}
\imath q q_{z} \, H_{n}^{(1)'}(q \tilde{d})   \\[0.6 em]  
 - \dfrac{n q_{z}}{\tilde{d}} \, H_{n}^{(1)}(q \tilde{d})   \\[0.7 em] 
 q^{2}  \, H_{n}^{(1)}(q \tilde{d})  \end{pmatrix} 
 \end{equation} \\
 
Recall that $\tilde{d} = g \, d$, with $g=2 \pi / \lambda$, where $d$ is the distance of the dipole to the SWCNT axis, and all dependence on the SWCNT radius appears in the reflection coefficients $ r_{\sigma,\sigma'} $.

\subsection{Propagating surface plasmon modes supported by a SWCNT}
The dispersion relation of surface plasmon modes can be obtained from the poles of the reflection coefficients, given by the zeroes of the denominator $D$ in Eqs.\eqref{E:reflection-coeff}. For small SWCNT radius, the relevant mode has $n=0$. Therefore, $N$ and $M$ waves decouple and the pole appears in the $r_{NN}^{n=0}$ coefficient, which is:
\begin{equation}
r_{NN}^{n=0} = - \frac{\pi \alpha \, q^{2}  \tilde{R} \, \left[J_{0}(q \tilde{R})\right]^2}{1 + \pi \alpha \, q^{2}  \tilde{R} \, J_{0}(q \tilde{R}) H_{0}^{(1)}(q \tilde{R}) } 
\end{equation}

Therefore the condition for existence of SPP is:
\begin{equation}
 q_{p}^{2}  \, J_{0}(q_{p} \tilde{R}) H_{0}^{(1)}(q_{p} \tilde{R}) = - \frac{1}{\pi \alpha \,  \tilde{R} }
\end{equation}

Recall that $q_{p}=\sqrt{1-q_{pz}^{2}}$. For very confined modes, $q_{pz}>>1$, so $q_{p}\approx \imath q_{pz}$ is a complex number with a large positive imaginary part. Using $K_{n}(z) = (\pi/2) \imath^{n+1} H_{n}^{(1)}(\imath z)$ and $I_{n}(z) = (\pi/2) \imath^{n} J_{n}(\imath z)$, we obtain the approximation, valid for $c \rightarrow \infty$, 

\begin{equation}
 q_{p}^{2}  \, I_{0}(q_{pz} \tilde{R}) K_{0}(q_{pz} \tilde{R}) =  \frac{\imath}{2 \alpha \,  \tilde{R} }
\end{equation}
which coincides with the result obtained within the quasi-static approximation.

In both cases, it is possible to write down the equations such that the left hand side depends on $x = q_{pz} \tilde{R}$ and the right hand side does not. For instance, using the quasi-static approximation, 
\begin{equation}
x^{2}  \, I_{0}(x) K_{0}(x) =  \frac{\imath  c \tilde{R}  }{ 4 \pi  \sigma }
\end{equation}
In order to estimate the plasmon wave vector, in the case of interest when the plasmon oscillates several times before it decays, we use the expression for $\sigma$ with $\tau = \infty$ ($\sigma = \imath \frac{e^2}{\pi \hbar^2} \frac{\mu_c }{\omega})$, 
arriving at
\begin{equation}
x^{2}  \, I_{0}(x) K_{0}(x) = \left(\frac{\omega }{ \omega_{0} }\right)^{2},
\end{equation}
where 
\begin{equation}
 \omega_{0} \equiv \frac{2 e}{\hbar}\, \sqrt{\frac{\mu_{c}}{R}}
\end{equation}
These expressions are simpler in both the limit for $x=q_{p} \tilde{R} << 1$, when $I_{0}(x) K_{0}(x)\approx - log(x)$
and $x=q_{p} \tilde{R} >>1$, when $I_{0}(x) K_{0}(x)\approx 1/2$.


\begin{thebibliography}{41}
\expandafter\ifx\csname natexlab\endcsname\relax\def\natexlab#1{#1}\fi
\expandafter\ifx\csname bibnamefont\endcsname\relax
  \def\bibnamefont#1{#1}\fi
\expandafter\ifx\csname bibfnamefont\endcsname\relax
  \def\bibfnamefont#1{#1}\fi
\expandafter\ifx\csname citenamefont\endcsname\relax
  \def\citenamefont#1{#1}\fi
\expandafter\ifx\csname url\endcsname\relax
  \def\url#1{\texttt{#1}}\fi
\expandafter\ifx\csname urlprefix\endcsname\relax\def\urlprefix{URL }\fi
\providecommand{\bibinfo}[2]{#2}
\providecommand{\eprint}[2][]{\url{#2}}

\bibitem[{\citenamefont{Chang et~al.}(2006)\citenamefont{Chang, S\"{o}rensen,
  Hemmer, and Lukin}}]{CSH06}
\bibinfo{author}{\bibfnamefont{D.~E.} \bibnamefont{Chang}},
  \bibinfo{author}{\bibfnamefont{A.~S.} \bibnamefont{S\"{o}rensen}},
  \bibinfo{author}{\bibfnamefont{P.~R.} \bibnamefont{Hemmer}},
  \bibnamefont{and} \bibinfo{author}{\bibfnamefont{M.~D.} \bibnamefont{Lukin}},
  \bibinfo{journal}{Phys. Rev. Lett.} \textbf{\bibinfo{volume}{97}},
  \bibinfo{pages}{053002} (\bibinfo{year}{2006}).

\bibitem[{\citenamefont{Akimov et~al.}(2007)\citenamefont{Akimov, Mukherjee,
  Yu, Chang, Zibrov, Hemmer, Park, and Lukin}}]{Akimov07}
\bibinfo{author}{\bibfnamefont{A.}~\bibnamefont{Akimov}},
  \bibinfo{author}{\bibfnamefont{A.}~\bibnamefont{Mukherjee}},
  \bibinfo{author}{\bibfnamefont{C.}~\bibnamefont{Yu}},
  \bibinfo{author}{\bibfnamefont{D.}~\bibnamefont{Chang}},
  \bibinfo{author}{\bibfnamefont{A.}~\bibnamefont{Zibrov}},
  \bibinfo{author}{\bibfnamefont{P.}~\bibnamefont{Hemmer}},
  \bibinfo{author}{\bibfnamefont{H.}~\bibnamefont{Park}}, \bibnamefont{and}
  \bibinfo{author}{\bibfnamefont{M.}~\bibnamefont{Lukin}},
  \bibinfo{journal}{Nature} \textbf{\bibinfo{volume}{450}},
  \bibinfo{pages}{402} (\bibinfo{year}{2007}).

\bibitem[{\citenamefont{Chang et~al.}(2007)\citenamefont{Chang, S\"{o}rensen,
  Demler, and Lukin}}]{CSD07}
\bibinfo{author}{\bibfnamefont{D.~E.} \bibnamefont{Chang}},
  \bibinfo{author}{\bibfnamefont{A.~S.} \bibnamefont{S\"{o}rensen}},
  \bibinfo{author}{\bibfnamefont{E.~A.} \bibnamefont{Demler}},
  \bibnamefont{and} \bibinfo{author}{\bibfnamefont{M.~D.} \bibnamefont{Lukin}},
  \bibinfo{journal}{Nat.\ Phys.} \textbf{\bibinfo{volume}{3}},
  \bibinfo{pages}{807} (\bibinfo{year}{2007}).

\bibitem[{\citenamefont{Nie and Emory}(1997)}]{NE97}
\bibinfo{author}{\bibfnamefont{S.}~\bibnamefont{Nie}} \bibnamefont{and}
  \bibinfo{author}{\bibfnamefont{S.~R.} \bibnamefont{Emory}},
  \bibinfo{journal}{Science} \textbf{\bibinfo{volume}{275}},
  \bibinfo{pages}{1102} (\bibinfo{year}{1997}).

\bibitem[{\citenamefont{Kundu et~al.}(2008)\citenamefont{Kundu, Le, Nordlander,
  and Halas}}]{KLN08}
\bibinfo{author}{\bibfnamefont{J.}~\bibnamefont{Kundu}},
  \bibinfo{author}{\bibfnamefont{F.}~\bibnamefont{Le}},
  \bibinfo{author}{\bibfnamefont{P.}~\bibnamefont{Nordlander}},
  \bibnamefont{and} \bibinfo{author}{\bibfnamefont{N.~J.} \bibnamefont{Halas}},
  \bibinfo{journal}{Chem.\ Phys.\ Lett.} \textbf{\bibinfo{volume}{452}},
  \bibinfo{pages}{115} (\bibinfo{year}{2008}).

\bibitem[{\citenamefont{Anger et~al.}(2006)\citenamefont{Anger, Bharadwaj, and
  Novotny}}]{Anger06}
\bibinfo{author}{\bibfnamefont{P.}~\bibnamefont{Anger}},
  \bibinfo{author}{\bibfnamefont{P.}~\bibnamefont{Bharadwaj}},
  \bibnamefont{and} \bibinfo{author}{\bibfnamefont{L.}~\bibnamefont{Novotny}},
  \bibinfo{journal}{Phys. Rev. Lett.} \textbf{\bibinfo{volume}{96}},
  \bibinfo{pages}{113002} (\bibinfo{year}{2006}).

\bibitem[{\citenamefont{K{\"u}hn et~al.}(2006)\citenamefont{K{\"u}hn,
  H{\aa}kanson, Rogobete, and Sandoghdar}}]{Kuhn06}
\bibinfo{author}{\bibfnamefont{S.}~\bibnamefont{K{\"u}hn}},
  \bibinfo{author}{\bibfnamefont{U.}~\bibnamefont{H{\aa}kanson}},
  \bibinfo{author}{\bibfnamefont{L.}~\bibnamefont{Rogobete}}, \bibnamefont{and}
  \bibinfo{author}{\bibfnamefont{V.}~\bibnamefont{Sandoghdar}},
  \bibinfo{journal}{Phys. Rev. Lett.} \textbf{\bibinfo{volume}{97}},
  \bibinfo{pages}{017402} (\bibinfo{year}{2006}).

\bibitem[{\citenamefont{Archambault et~al.}(2010)\citenamefont{Archambault,
  Marquier, Greffet, and Arnold}}]{AMG10}
\bibinfo{author}{\bibfnamefont{A.}~\bibnamefont{Archambault}},
  \bibinfo{author}{\bibfnamefont{F.}~\bibnamefont{Marquier}},
  \bibinfo{author}{\bibfnamefont{J.~J.} \bibnamefont{Greffet}},
  \bibnamefont{and} \bibinfo{author}{\bibfnamefont{C.}~\bibnamefont{Arnold}},
  \bibinfo{journal}{Phys. Rev. B} \textbf{\bibinfo{volume}{82}},
  \bibinfo{pages}{035411} (\bibinfo{year}{2010}).

\bibitem[{\citenamefont{Martin-Cano et~al.}(2010)\citenamefont{Martin-Cano,
  Mart\'{\i}n-Moreno, Garc\'{\i}a-Vidal, and Moreno}}]{Martin-Cano10}
\bibinfo{author}{\bibfnamefont{D.}~\bibnamefont{Martin-Cano}},
  \bibinfo{author}{\bibfnamefont{L.}~\bibnamefont{Mart\'{\i}n-Moreno}},
  \bibinfo{author}{\bibfnamefont{F.~J.} \bibnamefont{Garc\'{\i}a-Vidal}},
  \bibnamefont{and} \bibinfo{author}{\bibfnamefont{E.}~\bibnamefont{Moreno}},
  \bibinfo{journal}{Nano Letters} \textbf{\bibinfo{volume}{10}},
  \bibinfo{pages}{3129} (\bibinfo{year}{2010}).

\bibitem[{\citenamefont{Dzsotjan et~al.}(2010)\citenamefont{Dzsotjan,
  S\"{o}rensen, and Fleischhauer}}]{DSF10}
\bibinfo{author}{\bibfnamefont{D.}~\bibnamefont{Dzsotjan}},
  \bibinfo{author}{\bibfnamefont{A.~S.} \bibnamefont{S\"{o}rensen}},
  \bibnamefont{and}
  \bibinfo{author}{\bibfnamefont{M.}~\bibnamefont{Fleischhauer}},
  \bibinfo{journal}{Phys. Rev. B} \textbf{\bibinfo{volume}{82}},
  \bibinfo{pages}{075427} (\bibinfo{year}{2010}).

\bibitem[{\citenamefont{Gonzalez-Tudela
  et~al.}(2011)\citenamefont{Gonzalez-Tudela, Martin-Cano, Moreno,
  Mart\'{\i}n-Moreno, Tejedor, and Garc\'{\i}a-Vidal}}]{Gonzalez-Tudela11}
\bibinfo{author}{\bibfnamefont{A.}~\bibnamefont{Gonzalez-Tudela}},
  \bibinfo{author}{\bibfnamefont{D.}~\bibnamefont{Martin-Cano}},
  \bibinfo{author}{\bibfnamefont{E.}~\bibnamefont{Moreno}},
  \bibinfo{author}{\bibfnamefont{L.}~\bibnamefont{Mart\'{\i}n-Moreno}},
  \bibinfo{author}{\bibfnamefont{C.}~\bibnamefont{Tejedor}}, \bibnamefont{and}
  \bibinfo{author}{\bibfnamefont{F.~J.} \bibnamefont{Garc\'{\i}a-Vidal}},
  \bibinfo{journal}{Phys. Rev. Lett.} \textbf{\bibinfo{volume}{106}},
  \bibinfo{pages}{020501} (\bibinfo{year}{2011}).

\bibitem[{\citenamefont{Dzsotjan et~al.}(2011)\citenamefont{Dzsotjan, K\"astel,
  and Fleischhauer}}]{Dzsotjan11}
\bibinfo{author}{\bibfnamefont{D.}~\bibnamefont{Dzsotjan}},
  \bibinfo{author}{\bibfnamefont{J.}~\bibnamefont{K\"astel}}, \bibnamefont{and}
  \bibinfo{author}{\bibfnamefont{M.}~\bibnamefont{Fleischhauer}},
  \bibinfo{journal}{Phys. Rev. B} \textbf{\bibinfo{volume}{84}},
  \bibinfo{pages}{075419} (\bibinfo{year}{2011}).

\bibitem[{\citenamefont{Hanson}(2008)}]{Hanson08}
\bibinfo{author}{\bibfnamefont{G.~W.} \bibnamefont{Hanson}},
  \bibinfo{journal}{J. Appl. Phys.} \textbf{\bibinfo{volume}{103}},
  (\bibinfo{year}{2008}).

\bibitem[{\citenamefont{Jablan et~al.}(2009)\citenamefont{Jablan, Buljan, and
  Solja\ifmmode \check{c}\else \v{c}\fi{}i\ifmmode~\acute{c}\else
  \'{c}\fi{}}}]{Jablan09}
\bibinfo{author}{\bibfnamefont{M.}~\bibnamefont{Jablan}},
  \bibinfo{author}{\bibfnamefont{H.}~\bibnamefont{Buljan}}, \bibnamefont{and}
  \bibinfo{author}{\bibfnamefont{M.}~\bibnamefont{Solja\ifmmode \check{c}\else
  \v{c}\fi{}i\ifmmode~\acute{c}\else \'{c}\fi{}}}, \bibinfo{journal}{Phys. Rev.
  B} \textbf{\bibinfo{volume}{80}}, \bibinfo{pages}{245435}
  (\bibinfo{year}{2009}).

\bibitem[{\citenamefont{Castro~Neto et~al.}(2009)\citenamefont{Castro~Neto,
  Guinea, Peres, Novoselov, and Geim}}]{CastroNeto}
\bibinfo{author}{\bibfnamefont{A.~H.} \bibnamefont{Castro~Neto}},
  \bibinfo{author}{\bibfnamefont{F.}~\bibnamefont{Guinea}},
  \bibinfo{author}{\bibfnamefont{N.~M.~R.} \bibnamefont{Peres}},
  \bibinfo{author}{\bibfnamefont{K.~S.} \bibnamefont{Novoselov}},
  \bibnamefont{and} \bibinfo{author}{\bibfnamefont{A.~K.} \bibnamefont{Geim}},
  \bibinfo{journal}{Rev. Mod. Phys.} \textbf{\bibinfo{volume}{81}},
  \bibinfo{pages}{109} (\bibinfo{year}{2009}).

\bibitem[{\citenamefont{Fei et~al.}(2011)\citenamefont{Fei, Andreev, Bao,
  Zhang, S.~McLeod, Wang, Stewart, Zhao, Dominguez, Thiemens et~al.}}]{Fei11}
\bibinfo{author}{\bibfnamefont{Z.}~\bibnamefont{Fei}},
  \bibinfo{author}{\bibfnamefont{G.~O.} \bibnamefont{Andreev}},
  \bibinfo{author}{\bibfnamefont{W.}~\bibnamefont{Bao}},
  \bibinfo{author}{\bibfnamefont{L.~M.} \bibnamefont{Zhang}},
  \bibinfo{author}{\bibfnamefont{A.}~\bibnamefont{S.~McLeod}},
  \bibinfo{author}{\bibfnamefont{C.}~\bibnamefont{Wang}},
  \bibinfo{author}{\bibfnamefont{M.~K.} \bibnamefont{Stewart}},
  \bibinfo{author}{\bibfnamefont{Z.}~\bibnamefont{Zhao}},
  \bibinfo{author}{\bibfnamefont{G.}~\bibnamefont{Dominguez}},
  \bibinfo{author}{\bibfnamefont{M.}~\bibnamefont{Thiemens}},
  \bibnamefont{et~al.}, \bibinfo{journal}{Nano Letters}
  \textbf{\bibinfo{volume}{11}}, \bibinfo{pages}{4701} (\bibinfo{year}{2011}).

\bibitem[{\citenamefont{Chen et~al.}(2012)\citenamefont{Chen, Badioli,
  Alonso-Gonz\'alez, Thongrattanasiri, Huth, Osmond, Spasenovi\'c, Centeno,
  Pesquera, Godignon et~al.}}]{chen12}
\bibinfo{author}{\bibfnamefont{J.}~\bibnamefont{Chen}},
  \bibinfo{author}{\bibfnamefont{M.}~\bibnamefont{Badioli}},
  \bibinfo{author}{\bibfnamefont{P.}~\bibnamefont{Alonso-Gonz\'alez}},
  \bibinfo{author}{\bibfnamefont{S.}~\bibnamefont{Thongrattanasiri}},
  \bibinfo{author}{\bibfnamefont{F.}~\bibnamefont{Huth}},
  \bibinfo{author}{\bibfnamefont{J.}~\bibnamefont{Osmond}},
  \bibinfo{author}{\bibfnamefont{M.}~\bibnamefont{Spasenovi\'c}},
  \bibinfo{author}{\bibfnamefont{A.}~\bibnamefont{Centeno}},
  \bibinfo{author}{\bibfnamefont{A.}~\bibnamefont{Pesquera}},
  \bibinfo{author}{\bibfnamefont{P.}~\bibnamefont{Godignon}},
  \bibnamefont{et~al.}, \bibinfo{journal}{Nature}
  \textbf{\bibinfo{volume}{487}}, \bibinfo{pages}{77} (\bibinfo{year}{2012}).

\bibitem[{\citenamefont{Fei et~al.}(2012)\citenamefont{Fei, Rodin, Andreev,
  Bao, McLeod, Wagner, Zhang, Zhao, Thiemens, Dominguez et~al.}}]{FRA12}
\bibinfo{author}{\bibfnamefont{Z.}~\bibnamefont{Fei}},
  \bibinfo{author}{\bibfnamefont{A.~S.} \bibnamefont{Rodin}},
  \bibinfo{author}{\bibfnamefont{G.~O.} \bibnamefont{Andreev}},
  \bibinfo{author}{\bibfnamefont{W.}~\bibnamefont{Bao}},
  \bibinfo{author}{\bibfnamefont{A.~S.} \bibnamefont{McLeod}},
  \bibinfo{author}{\bibfnamefont{M.}~\bibnamefont{Wagner}},
  \bibinfo{author}{\bibfnamefont{L.~M.} \bibnamefont{Zhang}},
  \bibinfo{author}{\bibfnamefont{Z.}~\bibnamefont{Zhao}},
  \bibinfo{author}{\bibfnamefont{M.}~\bibnamefont{Thiemens}},
  \bibinfo{author}{\bibfnamefont{G.}~\bibnamefont{Dominguez}},
  \bibnamefont{et~al.}, \bibinfo{journal}{Nature}
  \textbf{\bibinfo{volume}{487}}, \bibinfo{pages}{82} (\bibinfo{year}{2012}).

\bibitem[{\citenamefont{Fang et~al.}(2013)\citenamefont{Fang, Thongrattanasiri,
  Schlather, Liu, Ma, Wang, Ajayan, Nordlander, Halas, and {Garc\'{\i}a de
  Abajo}}}]{Fang13}
\bibinfo{author}{\bibfnamefont{Z.}~\bibnamefont{Fang}},
  \bibinfo{author}{\bibfnamefont{S.}~\bibnamefont{Thongrattanasiri}},
  \bibinfo{author}{\bibfnamefont{A.}~\bibnamefont{Schlather}},
  \bibinfo{author}{\bibfnamefont{Z.}~\bibnamefont{Liu}},
  \bibinfo{author}{\bibfnamefont{L.}~\bibnamefont{Ma}},
  \bibinfo{author}{\bibfnamefont{Y.}~\bibnamefont{Wang}},
  \bibinfo{author}{\bibfnamefont{P.~M.} \bibnamefont{Ajayan}},
  \bibinfo{author}{\bibfnamefont{P.}~\bibnamefont{Nordlander}},
  \bibinfo{author}{\bibfnamefont{N.~J.} \bibnamefont{Halas}}, \bibnamefont{and}
  \bibinfo{author}{\bibfnamefont{F.~J.} \bibnamefont{{Garc\'{\i}a de Abajo}}},
  \bibinfo{journal}{ACS Nano} \textbf{\bibinfo{volume}{7}},
  \bibinfo{pages}{2388} (\bibinfo{year}{2013}).

\bibitem[{\citenamefont{Brar et~al.}(2013)\citenamefont{Brar, Jang, Sherrott,
  Lopez, and Atwater}}]{BJS13}
\bibinfo{author}{\bibfnamefont{V.~W.} \bibnamefont{Brar}},
  \bibinfo{author}{\bibfnamefont{M.~S.} \bibnamefont{Jang}},
  \bibinfo{author}{\bibfnamefont{M.}~\bibnamefont{Sherrott}},
  \bibinfo{author}{\bibfnamefont{J.~J.} \bibnamefont{Lopez}}, \bibnamefont{and}
  \bibinfo{author}{\bibfnamefont{H.~A.} \bibnamefont{Atwater}},
  \bibinfo{journal}{Nano\ Lett.} \textbf{\bibinfo{volume}{13}},
  \bibinfo{pages}{2541} (\bibinfo{year}{2013}).

\bibitem[{\citenamefont{Vakil and Engheta}(2011)}]{Vakil11}
\bibinfo{author}{\bibfnamefont{A.}~\bibnamefont{Vakil}} \bibnamefont{and}
  \bibinfo{author}{\bibfnamefont{N.}~\bibnamefont{Engheta}},
  \bibinfo{journal}{Science} \textbf{\bibinfo{volume}{332}},
  \bibinfo{pages}{1291} (\bibinfo{year}{2011}).

\bibitem[{\citenamefont{Koppens et~al.}(2011)\citenamefont{Koppens, Chang, and
  {Garc\'{\i}a de Abajo}}}]{Koppens11}
\bibinfo{author}{\bibfnamefont{F.~H.~L.} \bibnamefont{Koppens}},
  \bibinfo{author}{\bibfnamefont{D.~E.} \bibnamefont{Chang}}, \bibnamefont{and}
  \bibinfo{author}{\bibfnamefont{F.~J.} \bibnamefont{{Garc\'{\i}a de Abajo}}},
  \bibinfo{journal}{Nano Letters} \textbf{\bibinfo{volume}{11}},
  \bibinfo{pages}{3370} (\bibinfo{year}{2011}).

\bibitem[{\citenamefont{Nikitin et~al.}(2011)\citenamefont{Nikitin, Guinea,
  Garc\'{\i}a-Vidal, and Mart\'{\i}n-Moreno}}]{Nikitin11}
\bibinfo{author}{\bibfnamefont{A.~Y.} \bibnamefont{Nikitin}},
  \bibinfo{author}{\bibfnamefont{F.}~\bibnamefont{Guinea}},
  \bibinfo{author}{\bibfnamefont{F.~J.} \bibnamefont{Garc\'{\i}a-Vidal}},
  \bibnamefont{and}
  \bibinfo{author}{\bibfnamefont{L.}~\bibnamefont{Mart\'{\i}n-Moreno}},
  \bibinfo{journal}{Phys. Rev. B} \textbf{\bibinfo{volume}{84}},
  \bibinfo{pages}{195446} (\bibinfo{year}{2011}).

\bibitem[{\citenamefont{{Garc\'{\i}a de Abajo}}(2014)}]{DeAbajo14}
\bibinfo{author}{\bibfnamefont{F.~J.} \bibnamefont{{Garc\'{\i}a de Abajo}}},
  \bibinfo{journal}{ACS Photonics} \textbf{\bibinfo{volume}{1}},
  \bibinfo{pages}{135} (\bibinfo{year}{2014}).

\bibitem[{\citenamefont{Manjavacas et~al.}(2013)\citenamefont{Manjavacas,
  Marchesin, Thongrattanasiri, Koval, Nordlander, S\'{a}nchez-Portal, and
  {Garc\'{\i}a de Abajo}}}]{Manjavacas13}
\bibinfo{author}{\bibfnamefont{A.}~\bibnamefont{Manjavacas}},
  \bibinfo{author}{\bibfnamefont{F.}~\bibnamefont{Marchesin}},
  \bibinfo{author}{\bibfnamefont{S.}~\bibnamefont{Thongrattanasiri}},
  \bibinfo{author}{\bibfnamefont{P.}~\bibnamefont{Koval}},
  \bibinfo{author}{\bibfnamefont{P.}~\bibnamefont{Nordlander}},
  \bibinfo{author}{\bibfnamefont{D.}~\bibnamefont{S\'{a}nchez-Portal}},
  \bibnamefont{and} \bibinfo{author}{\bibfnamefont{F.~J.}
  \bibnamefont{{Garc\'{\i}a de Abajo}}}, \bibinfo{journal}{ACS\ Nano}
  \textbf{\bibinfo{volume}{7}}, \bibinfo{pages}{3635} (\bibinfo{year}{2013}).

\bibitem[{\citenamefont{Soto~Lamata
  et~al.}(10.1021/ph500377u)\citenamefont{Soto~Lamata, Alonso-González,
  Hillenbrand, and Nikitin}}]{ISL14}
\bibinfo{author}{\bibfnamefont{I.}~\bibnamefont{Soto~Lamata}},
  \bibinfo{author}{\bibfnamefont{P.}~\bibnamefont{Alonso-González}},
  \bibinfo{author}{\bibfnamefont{R.}~\bibnamefont{Hillenbrand}},
  \bibnamefont{and} \bibinfo{author}{\bibfnamefont{A.~Y.}
  \bibnamefont{Nikitin}}, \bibinfo{journal}{ACS Photonics}
  (\bibinfo{year}{10.1021/ph500377u}).

\bibitem[{\citenamefont{{St\'ephan} et~al.}(2002)\citenamefont{{St\'ephan},
  Taverna, Kociak, Suenaga, Henrard, and Colliex}}]{STK02}
\bibinfo{author}{\bibfnamefont{O.}~\bibnamefont{{St\'ephan}}},
  \bibinfo{author}{\bibfnamefont{D.}~\bibnamefont{Taverna}},
  \bibinfo{author}{\bibfnamefont{M.}~\bibnamefont{Kociak}},
  \bibinfo{author}{\bibfnamefont{K.}~\bibnamefont{Suenaga}},
  \bibinfo{author}{\bibfnamefont{L.}~\bibnamefont{Henrard}}, \bibnamefont{and}
  \bibinfo{author}{\bibfnamefont{C.}~\bibnamefont{Colliex}},
  \bibinfo{journal}{Phys. Rev. B} \textbf{\bibinfo{volume}{66}},
  \bibinfo{pages}{155422} (\bibinfo{year}{2002}).

\bibitem[{\citenamefont{Guo et~al.}(2004)\citenamefont{Guo, Chu, Wang, and
  Duan}}]{GCW04}
\bibinfo{author}{\bibfnamefont{G.~Y.} \bibnamefont{Guo}},
  \bibinfo{author}{\bibfnamefont{K.~C.} \bibnamefont{Chu}},
  \bibinfo{author}{\bibfnamefont{D.}~\bibnamefont{Wang}}, \bibnamefont{and}
  \bibinfo{author}{\bibfnamefont{C.}~\bibnamefont{Duan}},
  \bibinfo{journal}{Phys. Rev. B} \textbf{\bibinfo{volume}{69}},
  \bibinfo{pages}{205416} (\bibinfo{year}{2004}).

\bibitem[{\citenamefont{Swathi and Sebastian}(2010)}]{SS10}
\bibinfo{author}{\bibfnamefont{R.~S.} \bibnamefont{Swathi}} \bibnamefont{and}
  \bibinfo{author}{\bibfnamefont{K.~L.} \bibnamefont{Sebastian}},
  \bibinfo{journal}{J.\ Phys.\ Chem.} \textbf{\bibinfo{volume}{132}},
  \bibinfo{pages}{104502} (\bibinfo{year}{2010}).

\bibitem[{\citenamefont{Ma et~al.}(2009)\citenamefont{Ma, Wang, and
  Wang}}]{Ma09}
\bibinfo{author}{\bibfnamefont{J.}~\bibnamefont{Ma}},
  \bibinfo{author}{\bibfnamefont{J.~N.} \bibnamefont{Wang}}, \bibnamefont{and}
  \bibinfo{author}{\bibfnamefont{X.~X.} \bibnamefont{Wang}},
  \bibinfo{journal}{J. Mater. Chem.} \textbf{\bibinfo{volume}{19}},
  \bibinfo{pages}{3033} (\bibinfo{year}{2009}).

\bibitem[{\citenamefont{Sfeir et~al.}(2006)\citenamefont{Sfeir, Beetz, Wang,
  Huang, Huang, Huang, Hone, O'Brien, Misewich, Heinz et~al.}}]{SEW06}
\bibinfo{author}{\bibfnamefont{M.~Y.} \bibnamefont{Sfeir}},
  \bibinfo{author}{\bibfnamefont{T.}~\bibnamefont{Beetz}},
  \bibinfo{author}{\bibfnamefont{F.}~\bibnamefont{Wang}},
  \bibinfo{author}{\bibfnamefont{L.}~\bibnamefont{Huang}},
  \bibinfo{author}{\bibfnamefont{X.~M.~H.} \bibnamefont{Huang}},
  \bibinfo{author}{\bibfnamefont{M.}~\bibnamefont{Huang}},
  \bibinfo{author}{\bibfnamefont{J.}~\bibnamefont{Hone}},
  \bibinfo{author}{\bibfnamefont{S.}~\bibnamefont{O'Brien}},
  \bibinfo{author}{\bibfnamefont{J.~A.} \bibnamefont{Misewich}},
  \bibinfo{author}{\bibfnamefont{T.~F.} \bibnamefont{Heinz}},
  \bibnamefont{et~al.}, \bibinfo{journal}{Science}
  \textbf{\bibinfo{volume}{312}}, \bibinfo{pages}{554} (\bibinfo{year}{2006}).

\bibitem[{\citenamefont{Blancon et~al.}(2013)\citenamefont{Blancon, Paillet,
  Tran, Than, Guebrou, Ayari, {San Miguel}, Phan, Zahab, Sauvajo
  et~al.}}]{BPT13}
\bibinfo{author}{\bibfnamefont{J.-C.} \bibnamefont{Blancon}},
  \bibinfo{author}{\bibfnamefont{M.}~\bibnamefont{Paillet}},
  \bibinfo{author}{\bibfnamefont{H.~N.} \bibnamefont{Tran}},
  \bibinfo{author}{\bibfnamefont{X.~T.} \bibnamefont{Than}},
  \bibinfo{author}{\bibfnamefont{S.~A.} \bibnamefont{Guebrou}},
  \bibinfo{author}{\bibfnamefont{A.}~\bibnamefont{Ayari}},
  \bibinfo{author}{\bibfnamefont{A.}~\bibnamefont{{San Miguel}}},
  \bibinfo{author}{\bibfnamefont{N.-M.} \bibnamefont{Phan}},
  \bibinfo{author}{\bibfnamefont{A.-A.} \bibnamefont{Zahab}},
  \bibinfo{author}{\bibfnamefont{J.-L.} \bibnamefont{Sauvajo}},
  \bibnamefont{et~al.}, \bibinfo{journal}{Nat.\ Commun.}
  \textbf{\bibinfo{volume}{4}}, \bibinfo{pages}{2542} (\bibinfo{year}{2013}).

\bibitem[{\citenamefont{Liu et~al.}()\citenamefont{Liu, Hong, Choi, Jin, Capaz,
  Kim, Aloni, Wang, Bai, Louie et~al.}}]{LHC14}
\bibinfo{author}{\bibfnamefont{K.}~\bibnamefont{Liu}},
  \bibinfo{author}{\bibfnamefont{X.}~\bibnamefont{Hong}},
  \bibinfo{author}{\bibfnamefont{S.}~\bibnamefont{Choi}},
  \bibinfo{author}{\bibfnamefont{C.}~\bibnamefont{Jin}},
  \bibinfo{author}{\bibfnamefont{R.~B.} \bibnamefont{Capaz}},
  \bibinfo{author}{\bibfnamefont{J.}~\bibnamefont{Kim}},
  \bibinfo{author}{\bibfnamefont{S.}~\bibnamefont{Aloni}},
  \bibinfo{author}{\bibfnamefont{W.}~\bibnamefont{Wang}},
  \bibinfo{author}{\bibfnamefont{X.}~\bibnamefont{Bai}},
  \bibinfo{author}{\bibfnamefont{S.~G.} \bibnamefont{Louie}},
  \bibnamefont{et~al.}, \eprint{cond-mat/1311.3328}.

\bibitem[{tbp()}]{tbp}
\bibinfo{note}{L. Mart\'{\i}n-Moreno, F. J. Garc\'{\i}a de Abajo, and F. J. Garc\'{\i}a-Vidal, to be published}.

\bibitem[{\citenamefont{Longe and Bose}(1993)}]{Longe93}
\bibinfo{author}{\bibfnamefont{P.}~\bibnamefont{Longe}} \bibnamefont{and}
  \bibinfo{author}{\bibfnamefont{S.~M.} \bibnamefont{Bose}},
  \bibinfo{journal}{Phys. Rev. B} \textbf{\bibinfo{volume}{48}},
  \bibinfo{pages}{18239} (\bibinfo{year}{1993}).

\bibitem[{\citenamefont{Yannouleas et~al.}(1996)\citenamefont{Yannouleas,
  Bogachek, and Landman}}]{Yannouleas96}
\bibinfo{author}{\bibfnamefont{C.}~\bibnamefont{Yannouleas}},
  \bibinfo{author}{\bibfnamefont{E.~N.} \bibnamefont{Bogachek}},
  \bibnamefont{and} \bibinfo{author}{\bibfnamefont{U.}~\bibnamefont{Landman}},
  \bibinfo{journal}{Phys. Rev. B} \textbf{\bibinfo{volume}{53}},
  \bibinfo{pages}{10225} (\bibinfo{year}{1996}).

\bibitem[{\citenamefont{Jiang}(1996)}]{Jiang96}
\bibinfo{author}{\bibfnamefont{X.}~\bibnamefont{Jiang}},
  \bibinfo{journal}{Phys. Rev. B} \textbf{\bibinfo{volume}{54}},
  \bibinfo{pages}{13487} (\bibinfo{year}{1996}).

\bibitem[{\citenamefont{Novotny and Hecht}(2012)}]{Novotny}
\bibinfo{author}{\bibfnamefont{L.}~\bibnamefont{Novotny}} \bibnamefont{and}
  \bibinfo{author}{\bibfnamefont{B.}~\bibnamefont{Hecht}},
  \emph{\bibinfo{title}{Principles of Nano-optics}}
  (\bibinfo{publisher}{Cambridge University Press}, \bibinfo{year}{2012}).

\bibitem[{\citenamefont{Jackson}(1999)}]{J99}
\bibinfo{author}{\bibfnamefont{J.~D.} \bibnamefont{Jackson}},
  \emph{\bibinfo{title}{Classical Electrodynamics}}
  (\bibinfo{publisher}{Wiley}, \bibinfo{address}{New York},
  \bibinfo{year}{1999}).

\bibitem[{\citenamefont{Abramowitz and Stegun}(1972)}]{AS1972}
\bibinfo{author}{\bibfnamefont{M.}~\bibnamefont{Abramowitz}} \bibnamefont{and}
  \bibinfo{author}{\bibfnamefont{I.~A.} \bibnamefont{Stegun}},
  \emph{\bibinfo{title}{Handbook of Mathematical Functions}}
  (\bibinfo{publisher}{Dover}, \bibinfo{address}{New York},
  \bibinfo{year}{1972}).

\bibitem{Abajo14} F. J. {Garc\'{\i}a de Abajo}, ACS Photonics, 1, 135 (2014).
\bibitem{Thon12} S. Thongrattanasiri, A. Manjavacas, and F. J. {Garc\'{\i}a de Abajo}, ACS Nano 6, 1766 (2012).
\bibitem{J99} J. D. Jackson, {\it Classical Electrodynamics}, (Wiley, New York, 1999).
\bibitem{AS1972} M. Abramowitz and I. A. Stegun, {\it Handbook of Mathematical Functions}, (Dover, New York, 1972).
\bibitem{Chen} Chen-To Tai, {\it Dyadic Green Functions in Electromagnetic Theory}, IEEE Press Series on Electromagnetic Waves (1994).

\end{thebibliography}

\end{document}